# Sparse Signal Reconstruction via Iterative Support Detection*

Yilun Wang† Wotao Yin‡

**Abstract**

We present a novel sparse signal reconstruction method "ISD", aiming to achieve fast reconstruction and a reduced requirement on the number of measurements compared to the classical $\ell_1$ minimization approach. ISD addresses failed reconstructions of $\ell_1$ minimization due to insufficient measurements. It estimates a support set $I$ from a current reconstruction and obtains a new reconstruction by solving the minimization problem $\min\{\sum_{i\notin I}|x_i| : Ax = b\}$, and it iterates these two steps for a small number of times. ISD differs from the orthogonal matching pursuit (OMP) method, as well as its variants, because (i) the index set $I$ in ISD is not necessarily nested or increasing and (ii) the minimization problem above updates all the components of $x$ at the same time. We generalize the *Null Space Property* to *Truncated Null Space Property* and present our analysis of ISD based on the latter.

We introduce an efficient implementation of ISD, called threshold–ISD, for recovering signals with fast decaying distributions of nonzeros from compressive sensing measurements. Numerical experiments show that threshold–ISD has significant advantages over the classical $\ell_1$ minimization approach, as well as two state–of–the–art algorithms: the iterative reweighted $\ell_1$ minimization algorithm (IRL1) and the iterative reweighted least–squares algorithm (IRLS).

MATLAB code is available for download from http://www.caam.rice.edu/~optimization/L1/ISD/.

**Key words.** compressed sensing, l1 minimization, iterative support detection, basis pursuit.
**AMS subject classifications.** 68U10, 65K10, 90C25, 90C51.

# Contents

*This research was supported in part by NSF CAREER Award DMS-07-48839, ONR Grant N00014-08-1-1101, the U. S. Army Research Laboratory and the U. S. Army Research Office grant W911NF-09-1-0383, and an Alfred P. Sloan Research Fellowship.

†School of Civil and Environmental Engineering, Cornell University, Ithaca, New York, 14853, U.S.A. (yilun.wang@gmail.com). The author's part of work was done when he was a doctoral student at Rice University.

‡Department of Computational and Applied Mathematics, Rice University, Houston, Texas, 77005, U.S.A. (wotao.yin@rice.edu).

# 1 Introduction and Contributions

Brought to the research forefront by Donoho [13] and Candes, Romberg, and Tao [4], compressive sensing (CS) reconstructs a sparse unknown signal from a small set of linear projections. Let $\bar{x} \in \mathbb{R}^n$ denote a $k$-sparse[1] unknown signal and $b := A\bar{x} \in \mathbb{R}^m$ represent a set of $m$ linear projections of $\bar{x}$. The optimization problem

$$(\mathrm{P}_{\ell_0}) \quad \min_x \|x\|_0 \quad \text{s.t.} \quad Ax = b, \tag{1}$$

where $\|x\|_0$ is defined as the number of nonzero components of $x$, can exactly reconstruct $\bar{x}$ from $O(k)$ random projections. (Throughout this paper, $\bar{x}$ is used to denote the true signal to reconstruct.) However, because $\|x\|_0$ is non-convex and combinatorial, $(\mathrm{P}_{\ell_0})$ is impractical for real applications. A practical alternative is the basis pursuit (BP) problem

$$(\mathrm{BP}) \qquad \min_x \|x\|_1 \quad \text{s.t.} \quad Ax = b, \tag{2}$$

$$\text{or} \qquad \min_x \; \|x\|_1 + \frac{1}{2\rho}\|Ax - b\|_2^2, \tag{3}$$

where (2) is used when $b$ contains little or no noise and, otherwise, (3) is used with a proper parameter $\rho > 0$. The BP problems have been known to yield sparse solutions under certain conditions (see [14, 11, 17] for explanations) and also have recent algorithms such as [3, 23, 20, 16, 15, 27, 30, 29]. It is shown in [6, 25] that, when $A$ is a Gaussian random or partial Fourier ensemble, BP returns a solution equal to $\bar{x}$ with high probability from $m = O(k\log(n/k))$ and $O(k\log(n)^4)$ linear measurements, respectively, which are much smaller than $n$. Compared to $(\mathrm{P}_{\ell_0})$, BP is much easier to solve but requires significantly more measurements.

We propose an iterative support detection method (abbreviated as ISD) that runs as fast as the best BP algorithms but requires significantly fewer measurements. ISD alternatively calls its two components: support detection and signal reconstruction. From an incorrect reconstruction, support detection identifies an index set $I$ containing some elements of $\mathrm{supp}(\bar{x}) = \{i : x_i \neq 0\}$, and signal reconstruction solves

$$(\text{Truncated BP}) \quad \min_x \; \|x_T\|_1 \quad \text{s.t.} \quad Ax = b, \tag{4}$$

where $T = I^C$ and $\|x_T\|_1 = \sum_{i \notin I} |x_i|$ (or solving a least-squares penalty version corresponding to (3)). Assuming a sparse original signal $\bar{x}$, if $I = \mathrm{supp}(\bar{x})$, then the solution of (4) is of course equal to $\bar{x}$. But this also happens if $I$ contains enough, not necessarily all, entries of $\mathrm{supp}(\bar{x})$. When $I$ does not have enough of $\mathrm{supp}(\bar{x})$ for an exact reconstruction, those entries of $\mathrm{supp}(\bar{x})$ in $I$ will help (4) return a better solution, which has a reduced error compared to the solution of (2). From this better solution, support detection will be able identify more entries in $\mathrm{supp}(\bar{x})$ and thus yield a better $I$. In this way, the two components of ISD work together to gradually recover $\mathrm{supp}(\bar{x})$ and improve the reconstruction. Given sufficient measurements, ISD can finally recover $\bar{x}$. Furthermore, exact reconstruction can happen even if $I$ includes a small number of the spurious indices out of $\mathrm{supp}(\bar{x})$. A simple demo in Section 2 below illustrates the above for a sparse Gaussian signal.

ISD requires the reliable support detection from inexact reconstructions, which must take advantages of the features and prior information about the true signal $\bar{x}$. In this paper, we focus on the sparse or compressible signals with components having a fast decaying distribution of nonzeros. For these signals, we perform support detection by thresholding the solution of (4) and call the corresponding ISD algorithm as *threshold–ISD*. We present different thresholding rules including a simple one given along with the demo in Section 2 and a more efficient one discussed in Subsection 4.1. The latter rule was used throughout our numerical experiments in Section 5.

[1] A $k$-sparse vector has no more than $k$ nonzero components.

To provide theoretical explanations for ISD, we analyze the model (4) based on a so-called *truncated null space property* of $A$, an extension of the null space property originally studied in [9] and later in [31, 32, 13, 10], which gives the widely used restricted isometry property [5] in certain cases. We establish sufficient conditions for (4) to return $\bar{x}$ exactly. When $\bar{x}$ is not exactly sparse, its exact reconstruction is generally impossible. We show an error bound between the solution of (4) and $\bar{x}$. Built upon these results for a single instance of (4), the following result for ISD is obtained: the chance for (4) to return a sparse signal $\bar{x}$ improves if in the new detections at each iteration, the true nonzeros are more than the false ones by a certain factor. These results are independent of specific support detection methods used for generating $I$. However, we yet to obtain a global convergence result for ISD.

While a recovery guarantee has not been obtained, numerical comparisons to state-of-the-art algorithms show that threshold–ISD runs very fast and requires very fewer measurements. Threshold–ISD calls YALL1 [30] with warm-start and a dynamic stopping rule to efficiently solve (4). As a result, the threshold–ISD time is comparable to the YALL1 time for solving the BP model. Threshold–ISD was compared to BP, the iteratively reweighted least squares algorithm [8] (IRLS) and the iteratively reweighted $\ell_1$ minimization algorithm [7] (IRL1) on various types of synthetic and real data. IRLS and IRL1 are known for their state-of-the-art reconstruction rates, on both noiseless and noisy measurements. Given the same number of measurements, threshold–ISD and IRLS returned better signals than IRL1, which is further better than BP. Comparing threadhold–ISD and IRLS, the former ran order-of-magnitude faster.

The rest of this paper is organized as follows. In Section 2, the algorithmic framework of ISD is given along with a simple demo. Sections 3 presents preliminary theoretical results. Section 4 and 5 study the details of threshold–ISD and present our numerical results, respectively. Section 6 is devoted to conclusions and discussions on future research.

## 2 Algorithmic Framework

We first present the algorithmic framework of ISD.

Input: $A$ and $b$

1. Set the iteration number $s \leftarrow 0$ and initialize the set of detected entries $I^{(s)} \leftarrow \emptyset$;
2. While the stopping condition is not met, do
   (a) $T^{(s)} \leftarrow (I^{(s)})^C := \{1, 2, \dots, n\} \setminus I^{(s)}$;
   (b) $x^{(s)} \leftarrow$ solve truncated BP (4) for $T = T^{(s)}$;
   (c) $I^{(s+1)} \leftarrow$ support detection using $x^{(s)}$ as the reference;
   (d) $s \leftarrow s + 1$.

Since $T^{(0)} = \{1, \dots, n\}$, (4) in Step (b) reduces to BP (2) in iteration 0.

Like *greedy algorithms* such as OMP [26], StOMP [12], and CoSaMP [24], ISD iteratively maintains an set $I$ of selected indices and updates $x$. However, ISD differs from greedy algorithms in how $I$ is grown and $x^{(s)}$ is updated. The index set $I$ in ISD is not necessarily nested or increasing over the iterations. This is more like CoSaMP, different from OMP and StOMP. At each iteration, after the index set $I$ is computed, ISD updates all the components of $x$ including both the detected and undetected ones at the same time. Both these differences are important since they allow ISD to reconstruct certain sparse signals that cannot be recovered by the existing greedy algorithms.

**A demo:** We generated a sparse signal $\bar{x}$ of length $n = 200$ with $k = 25$ nonzero numbers independently sampled from the standard Gaussian distribution and assigned to randomly chosen components of $\bar{x}$. We let $m = 60$, created a Gaussian random $m \times n$ matrix $A$, and set $b := A\bar{x}$. We implemented Step (c) by a threshold rule:

$$I^{(s+1)} \leftarrow \{i : |x_i^{(s)}| > \epsilon^{(s)}\}. \tag{5}$$

To keep it simple for now, we let

$$\epsilon^{(s)} := \|x^{(s)}\|_\infty / \beta^{(s+1)}. \tag{6}$$

with $\beta = 5$. Note that in Subsection 4.1 below, we will present a more reliable rule to determine $\epsilon^{(s)}$.

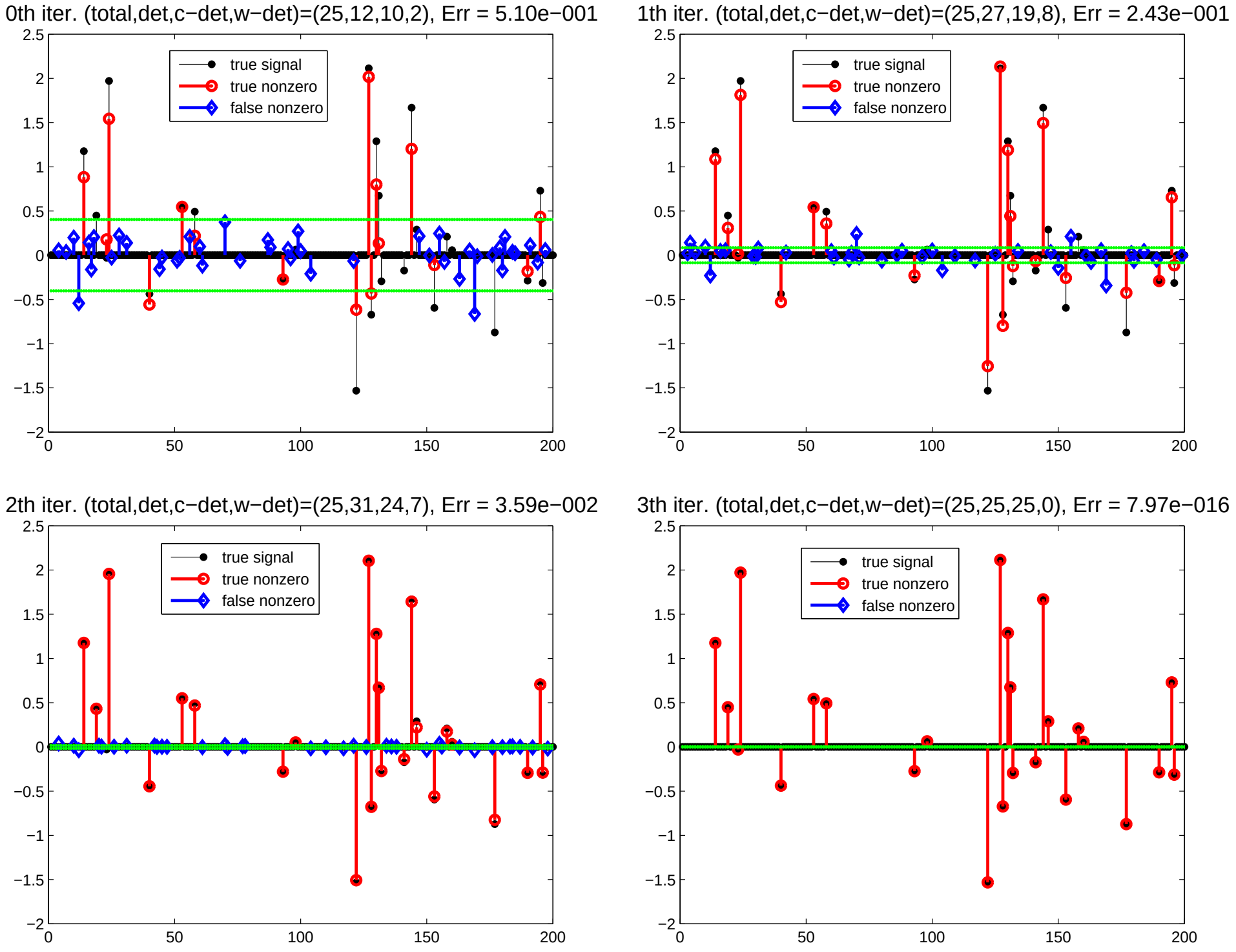

| | Itr $k$ | Nonzeros | | | | Relative Error |
|---|---|---|---|---|---|---|
| | | Total True | Detected | Correct | False | |
| BP | 0 | 25 | 12 | 10 | 2 | 5.10e-1 |
| | 1 | 25 | 27 | 19 | 8 | 2.43e-1 |
| | 2 | 25 | 31 | 24 | 7 | 3.59e-2 |
| | 3 | 25 | 25 | 25 | 0 | 7.97e-16 |

Figure 1: An ISD demo that recovers $\bar{x}$ with 25 nonzeros from 60 random Gaussian measurements.

With 200 dimensions, it is normally considered difficult to recover a signal with 25 nonzeros from merely 60 measurements (a 2.4x measurement-to-nonzero ratio), but ISD returns an exact reconstruction in merely four iterations. The solutions of the four iterations are depicted in the four subplots of Figure 1, where the components of $\bar{x}$ are marked by $\bullet$ and the nonzero components of $x^{(s)}$ are marked separately by $\circ$ and $\diamond$, standing for true and false nonzeros, respectively. The thresholds $\epsilon^{(s)}$ are shown as green lines. To measure the solution qualities, we give the quadruplet "(total, det, c–det, w–det)" and "Err" in the title of each subplot and in the table, which are defined as follows:

- (total, det, c-det, w-det):
  - total: the number of total nonzero components of the true signal $\bar{x}$.
  - det: the number of detected nonzero components, equal to $|I^{(s+1)}| = $ (c–det)+(w–det).

– c-det: the number of *correctly* detected nonzero components, i.e, $|I^{(s+1)} \cap \{i : \bar{x}_i \neq 0\}|$.
  – w-det: the number of *falsely* detected nonzero components, i.e, $|I^{(s+1)} \cap \{i : \bar{x}_i = 0\}|$.

- Err: the relative error $\|x^{(s)} - \bar{x}\|_2 / \|\bar{x}\|_2$.

From the upper left subplot, it is clear that $x^{(0)}$, which was the BP solution, contained a large number of false nonzeros and had a large relative error. However, most of its correct nonzero components were relatively large in magnitude (as a consequence of $\bar{x}$ having a relatively fast decaying distribution of nonzeros), the thresholding method (5) with the threshold $\epsilon^{(0)} = \|x^{(0)}\|_\infty/5$ detected 12 nonzeros, among which 10 were true nonzeros and 2 were not. In spite of the 2 false detections, the detection yielded $T^{(1)}$ that was good to let (4) return a much better solution $x^{(1)}$, depicted in the upper right subplot. This solution further allowed (5), now having the tighter threshold $\epsilon^{(1)} = \|x^{(1)}\|_\infty/5^2$, to yield 19 detected true nonzeros with 8 false detections. Noticeably, most of true nonzeros with large magnitude had been correctly detected. The next solution $x^{(2)}$, depicted in the bottom left subplot, became even better, which well matched the true signal $\bar{x}$ except for tiny false nonzero components. (5) detected 24 true nonzeros of $\bar{x}$ from $x^{(2)}$ and 7 false nonzeros, and $x^{(3)}$ had exactly the same nonzero components as $\bar{x}$, as well as an error almost as low as the double precision.

ISD is insensitive to a small number of false detections and has an attractive self-correction capacity. It is important to generate every $I$ from $x^{(s)}$ regardless what it was previously; otherwise, false detection would be trapped in $I$. In addition, the performance of ISD is invariant to small variations in the thresholding tolerance in (5). On the same set of data, we also tried to set $\beta = 3$ and $\beta = 1.5$ in (6), and obtained an exact reconstruction of $\bar{x}$ in 4 and 6 iterations, respectively.

# 3 Preliminary Theoretical Analysis

The preliminary theoretical results in this section explain under what conditions the truncated BP model (4) can successfully reconstruct $\bar{x}$, especially, from measurements that are not enough for BP. Most of the results are based on a property of the sensing matrix $A$ defined in Subsection 3.1. Focusing on the minimization problem (4), Subsections 3.2 and 3.3 study exact reconstruction conditions for sparse signals and reconstruction errors for compressible signals, respectively. Finally, Subsection 3.4 gives a sufficient condition for ISD to improve the chance of perfect reconstruction over its iteration. Note that this condition is not a recovery guarantee, which is yet to be found.

## 3.1 The Truncated Null Space Property

We start with introducing the *truncated null space property* ($t$-NSP), a generalization of the *null space property* (NSP). The NSP is used in slightly different forms and different names in [31, 32, 13, 9, 10]. We adopt the definition in [10]: a matrix $A \in \mathbb{R}^{m\times n}$ satisfies the NSP of order $L$ for $\gamma > 0$ if

$$\|\eta_S\|_1 \leq \gamma \|\eta_{S^C}\|_1 \tag{7}$$

holds for all index sets $S$ with $|S| \leq L$ and all $\eta \in \mathcal{N}(A)$, which is the null space of $A$. In (7) and the rest of this paper, $\eta_S \in \mathbb{R}^n$ denotes the subvector of $\eta$ consisting of $\eta_i$ for $i \in S \subset \{1, 2, \ldots, n\}$, and $S^C$ denotes the complement of $S$ with respect to $\{1, \ldots, n\}$.

With $\gamma < 1$, the NSP says that any nonzero vector $\eta$ in the null space of $A$ cannot have an $\ell_1$-mass concentrated on any set with $L$ or fewer elements. A sufficient exact reconstruction condition for BP is given in [10] based on the NSP: the true $k$–sparse signal $\bar{x}$ is the unique solution of BP if $A$ has the NSP of order $L \geq k$ and $0 < \gamma < 1$.

In order to analyze the minimization problem (4) with a truncated $\ell_1$–norm objective, we now generalize the NSP to the $t$-NSP.

**Definition 1.** *A matrix $A$ satisfies the $t$-NSP of order $L$ for $\gamma > 0$ and $0 < t \leq n$ if*

$$\|\eta_S\|_1 \leq \gamma \|\eta_{(T \cap S^C)}\|_1 \tag{8}$$

*holds for all sets $T \subset \{1, \ldots, n\}$ with $|T| = t$, all subsets $S \subset T$ with $|S| \leq L$, and all $\eta \in \mathcal{N}(A)$ — the null space of $A$.*

*For simplicity, we use $t$-NSP$(t, L, \gamma)$ to denote the $t$-NSP of order $L$ for $\gamma$ and $t$, and use $\bar{\gamma}$ to replace $\gamma$ and write $t$-NSP$(t, L, \bar{\gamma})$ if $\bar{\gamma}$ is the infimum of all the feasible $\gamma$ satisfying (8).*

Notice that when $t = n$, the $t$-NSP reduces to the NSP. Compared to the NSP, the inequality (8) in the $t$-NSP has an extra set limiter $T$ of size $t$. It is introduced to deal with $\|x_T\|_1$.

Clearly, for a given $A$ and $t$, $\bar{\gamma}$ is monotonic increasing in $L$. On the other hand, fixing $L$, $\bar{\gamma}$ is monotonically decreasing in $t$. If $\gamma$ is fixed, then the largest legitmate $L$ is monotonically increasing in $t$.

## 3.2 Sufficient Recovery Conditions of Truncated $\ell_1$ Minimization

We first analyze the model (4) and explain why it may require significantly fewer measurements than BP. Below we present a sufficient exact reconstruction condition, in which the requirement $\|\bar{x}\|_0 \leq L$ for BP is replaced by $\|\bar{x}_T\|_0 \leq L$.

**Theorem 3.1.** *Let $\bar{x}$ be a given vector and $T$ be a given index set satisfying $T \cap \operatorname{supp}(\bar{x}) \neq \emptyset$. Assume that a matrix $A$ satisfies $t$-NSP$(t, L, \bar{\gamma})$ for $t = |T|$. If $\|\bar{x}_T\|_0 \leq L$ and $\bar{\gamma} < 1$, then $\bar{x}$ is the unique minimizer of (4) for $b := A\bar{x}$.*

*Proof.* The true signal $\bar{x}$ uniquely solves (4) if and only if

$$\|\bar{x}_T + v_T\|_1 > \|\bar{x}_T\|_1, \ \forall v \in \mathcal{N}(A), v \neq \mathbf{0}. \tag{9}$$

Let $S := T \cap \operatorname{supp}(\bar{x})$. Since $\|\bar{x}_S\|_1 = \|\bar{x}_T\|_1$, we have

$$\begin{aligned} \|\bar{x}_T + v_T\|_1 &= \|\bar{x}_S + v_S\|_1 + \|\mathbf{0} + v_{T \cap S^C}\|_1 \\ &= \underbrace{(\|\bar{x}_S + v_S\|_1 - \|\bar{x}_S\|_1 + \|v_S\|_1)}_{\geq \mathbf{0}} + \|\bar{x}_T\|_1 \\ &+ \ (\|v_{T \cap S^C}\|_1 - \|v_S\|_1). \end{aligned}$$

Therefore, having $\|v_S\|_1 < \|v_{T \cap S^C}\|_1$ is sufficient for (9).

If $\|\bar{x}_T\|_0 \leq L$, then $|S| \leq L$. According to the definition of $t$-NSP$(|T|, L, \bar{\gamma})$, it holds that $\|v_S\|_1 \leq \bar{\gamma}\|v_{T \cap S^C}\|_1 < \|v_{T \cap S^C}\|_1$. □

The assumption $T \cap \operatorname{supp}(\bar{x}) \neq \emptyset$ in Theorem 3.1 is not essential because otherwise, $\bar{x}$ is a trivial solution of (4). In addition if $A_{T^C}$ has independent columns, then $\bar{x}$ is the unique solution. We note that $t$-NSP$(|T|, L, \bar{\gamma})$ is more strict than what is needed when $T$ is given because (8) is required to hold for all $T$ with $|T| = t$.

The following lemma states that the $t$-NSP is satisfied by Gaussian matrices of appropriate sizes. Our proof is inspired by the work [33].

**Lemma 3.1.** *Let $m < n$. Assume that $A \in \mathbb{R}^{m \times n}$ is either a standard Gaussian matrix (i.e., one with i.i.d. standard normal entries) or there exists a standard Gaussian matrix $B \in \mathbb{R}^{n \times (n-m)}$ such that $AB = \mathbf{0}$. Given an index set $T$, with probability greater than $1 - e^{-c_0(n-m)}$, the matrix $A$ satisfies $t$-NSP$(t, L, \gamma)$ for $t = |T|$ and $\gamma = \frac{\sqrt{L}}{2\sqrt{k(d)} - \sqrt{L}}$, where*

$$k(d) := c\frac{m - d}{1 + \log(\frac{n-d}{m-d})}, \tag{10}$$

*$d = n - |T|$, and $c_0, c > 0$ are absolute constants independent of the dimensions $m$, $n$ and $d$.*

In ISD, $d$ equals the number of detected entries (including both correct and false detections). $d$ determines $k(d)$ and in turn the $t$-NSP parameter $\gamma$. Since the sufficient condition of Theorem 3.1 requires $\bar{\gamma} < 1$ (i.e., there exists a $\gamma < 1$), we see that support detection affects the chance of recovery by (4). Because $k(d)$ plays a pivoting role here, we analyze its formula (10) at the end of this subsection. Also, we note that the $\gamma$ given in Lemma 3.1 is not necessarily tight.

*Proof.* Let the columns of $B$ span $\mathcal{N}(A)$, i.e., $B \in \mathbb{R}^{(n-m)\times m}$ and $AB = \mathbf{0}$, and $P_T$ refer to projection to the coordinates $T$. Then, $\Lambda = \{v_T : v \in \mathcal{N}(A)\} = \{(P_T B)w : w \in \mathbb{R}^{n-m}\}$ is a randomly drawn subspace in $\mathbb{R}^{|T|}$ with dimensions up to $(n-m)$. Kashin-Garnaev-Gluskin's result [22, 18] states that for any $p < q$, with

$$\text{probability} \geq 1 - e^{-c_0 p},$$

a randomly drawn $p$-dimensional subspace $V_p \in \mathbb{R}^q$ satisfies

$$\frac{\|z\|_1}{\|z\|_2} \geq \frac{c_1\sqrt{q-p}}{\sqrt{1+\log(q/(q-p))}}, \ \forall z \in V_p, z \neq \mathbf{0},$$

where $c_0$ and $c_1$ are independent of the dimensions. Applying this result with $q := |T| = n-d$ and $p := n-m$, we obtain

$$\frac{\|v_T\|_1}{\|v_T\|_2} \geq \frac{c_1\sqrt{(n-d)-(n-m)}}{\sqrt{1+\log((n-d)/((n-d)-(n-m))}}, \ \forall v \in \mathcal{N}(A), \ v \neq \mathbf{0}.$$

or

$$\frac{\|v_T\|_1}{\|v_T\|_2} \geq \frac{c_1\sqrt{m-d}}{\sqrt{1+\log\frac{n-d}{m-d}}}, \ \forall v \in \mathcal{N}(A), \ v \neq \mathbf{0}.$$

Let $k(d)$ be defined in (10) where $c = \frac{c_1^2}{4}$. For all $S \subset T$ with $|S| \leq \ L$ we have $\sqrt{k(d)}\|v_T\|_2 \leq \frac{1}{2}\|v_T\|_1$ and thus

$$\|v_S\|_1 \leq \sqrt{|S|}\|v_S\|_2 \leq \frac{\sqrt{L}}{\sqrt{k(d)}}\sqrt{k(d)}\|v_S\|_2 \leq \frac{\sqrt{L}}{\sqrt{k(d)}}\sqrt{k(d)}\|v_T\|_2 \leq \frac{\sqrt{L}}{2\sqrt{k(d)}}\|v_T\|_1,$$

or equivalently

$$\|v_S\|_1 \leq \gamma\|v_{T\cap S^C}\|_1,$$

where $\gamma = \frac{\sqrt{L}}{2\sqrt{k(d)}-\sqrt{L}}$. The lemma follows from the definition of the $t$-NSP.

□

Theorem 3.1 and Lemma 3.1 lead to the following theorem.

**Theorem 3.2.** *Let $\bar{x} \in \mathbb{R}^n$ and $T$ be given such that $T \cap \operatorname{supp}(\bar{x}) \neq \emptyset$. Let $m < n$ and $A \in \mathbb{R}^{m\times n}$ be given as in Lemma 3.1. Then, with probability greater than $1 - e^{-c_0(n-m)}$, the true signal $\bar{x}$ is the unique solution of (4) for $b := A\bar{x}$ if*

$$\|\bar{x}_T\|_0 < k(d), \tag{11}$$

*where $k(d)$ is defined in (10), $d = n - t = n - |T|$, and $c_0, c > 0$ are absolute constants independent of the dimensions $m$, $n$, and $d$.*

*Furthermore, let $d_c = |I \cap \operatorname{supp}(\bar{t})|$ denote the number of correct detections. Then, (11) is equivalent to*

$$\|\bar{x}\|_0 < k(d) + d_c. \tag{12}$$

*Proof.* In Lemma 3.1, let $L = \|\bar{x}_T\|_0$ and if $L < k(d)$, then $\gamma < 1$. Then, the first result follows from Theorem 3.1. (11) and (12) are equivalent since $\|\bar{x}\|_0 = \|\bar{x}_T\|_0 + d_c$. □

Note that when $d = 0$, condition (11) reduces to the existing result for BP: for the same vector $\bar{x}$ and $A$ given in Theorem 3.2 above, with probability greater than $1 - e^{-c_0(n-m)}$, $\bar{x}$ is the unique solution of (4) for $b := A\bar{x}$ if $\|\bar{x}\|_0 \leq cm(1 + \log(n/m))^{-1}$, namely, the inequality (11) holds for $d = 0$.

In light of (12), to compare BP with truncated BP, we shall compare $k(0)$ with $k(d) + d_c$. Below we argue that if there are enough correct detections (i.e., $d_c/d$ is sufficiently large), then we get $k(0) < k(d) + d_c$ and it is easier for truncated BP to recover $\bar{x}$ than BP. To see this, we start with

$$k(0) = k(d) - \int_0^d k'(d) \tag{13}$$

and study $k'(d)$. Because (4) is equivalent to a linear program which has a solution with no more than $m$ nonzeros, we naturally assume $d < m$. Then, we obtain

$$k'(d) := -c\left(\frac{1}{1+\log\left(\frac{n-d}{m-d}\right)} + \frac{n-m}{\left(1+\log\left(\frac{n-d}{m-d}\right)\right)^2 (n-d)}\right) < 0. \tag{14}$$

On the other hand, we have $-1 < k'(d)$ for the following reasons. First, it is well-known that universal stable reconstruction (by any methods) requires $\|\bar{x}\|_0 < m/2$, which we now assume. Second, when BP fails (which is the case we are interested in), Theorem 3.2 gives us $k(0) \le \|\bar{x}\|_0$. Therefore, we have $k(0) < m/2$ or

$$\frac{c}{1+\log(\frac{n}{m})} < \frac{1}{2}. \tag{15}$$

Plugging (15) into (14) one can deduce $-1 < k'(d) < 0$, together with (13) which gives

$$k(0) = k(d) - \int_0^d k'(d) < k(d) + d. \tag{16}$$

(16) means that if $d = d_c$ (i.e., there is no false detection), truncated BP does better than BP. For practical reasons, we do not always have $d = d_c$, but on the other hand, one does not push $\|\bar{x}\|_0$ to its limit at $m/2$. Consequently, (15) is very conservative. In fact, it is safe to assume $\frac{c}{1+\log(\frac{n}{m})} < 1/4$, which gives $-1/2 < k'(d) < 0$ and thus

$$k(0) = k(d) - \int_0^d k'(d) < k(d) + \frac{1}{2}d. \tag{17}$$

Comparing the right-hand sides of (12) and (17), we can see that as long as $d_c \ge (1/2)d$ (i.e., at least half of the detections are correct), then truncated BP is more likely to recover $\bar{x}$ than BP. It is easy to extend this analysis to the iterations of ISD as follows. Since $k(d)$ reduces at rate slower than $1/2$, $k(d)+d_c$ will increase as long as $d_c$ grows faster in terms of $\Delta d_c/\Delta d > 1/2$ between iterations. Finally, we can also observe that the higher the sample ratio $m/n$, the smaller $|k'(d)|$. Hence, the sufficient reconstruction condition becomes even easier to satisfy.

## 3.3 Stability of Truncated $\ell_1$ Minimization

Because many practical signals are not exactly sparse, we study the reconstruction error of (4) applied to general signals, which is expressed in the best $L$–term approximation error of $\bar{x}$:

$$\sigma_L(\bar{x})_1 := \inf\{\|\bar{x} - x\|_1 : \|x\|_0 \le L, x \in \mathbb{R}^{\dim(\bar{x})}\}.$$

For a signal $\bar{x}$ with a fast decaying tail (in terms of the distribution of its entries), this error is much smaller than $\|\bar{x}\|_1$. Theorem 3.3 below states that under certain conditions on $A$, (4) returns a solution with an $\ell_1$-error bounded by $\sigma_L(\bar{x})$ up to a constant factor depending only on $|T|$. The theorem needs the following lemma, which is an extension of Lemma 4.2 in [10].

**Lemma 3.2.** *Consider problem (4) with a given $T$, and let $z, z' \in \mathcal{F}(b)$. Assume that $A$ satisfies t-NSP$(t, L, \bar{\gamma})$, where $t = |T|$ and $\bar{\gamma} < 1$. Let $S \subset T$ be the set of indices corresponding to the largest $L$ entries in $z_T$. We have*

$$\|(z - z')_{T\cap S^C}\|_1 \le \frac{1}{1-\bar{\gamma}}\left(\|z'_T\|_1 - \|z_T\|_1 + 2\sigma_L(z_T)_1\right), \tag{18}$$

*where $\sigma_L(z_T)_1$ is the best $L$-term approximation error of $z_T$.*

*Proof.* We have $\|z_{T\cap S^C}\|_1 = \sigma_L(z_T)_1$ and

$$\begin{aligned}
\|(z'-z)_{T\cap S^C}\|_1 &\leq \|z'_{T\cap S^C}\|_1 + \|z_{T\cap S^C}\|_1 \\
&= \|z'_T\|_1 - \|z'_S\|_1 + \sigma_L(z_T)_1 \\
&= \|z_T\|_1 + \|z'_T\|_1 - \|z_T\|_1 - \|z'_S\|_1 + \sigma_L(z_T)_1 \\
&= \|z_S\|_1 - \|z'_S\|_1 + \|z'_T\|_1 - \|z_T\|_1 + 2\sigma_L(z_T)_1 \\
&\leq \|(z-z')_S\|_1 + \|z'_T\|_1 - \|z_T\|_1 + 2\sigma_L(z_T)_1.
\end{aligned}$$

Equation (18) follows from the above inequality and the definition of $t$-NSP$(t, L, \bar{\gamma})$, which says

$$\|(z'-z)_S\|_1 \leq \bar{\gamma}(\|(z'-z)_{T\cap S^C}\|_1). \tag{19}$$

□

Lemma 3.2 leads to the following theorem.

**Theorem 3.3.** *Consider problem (4) for a given $T$. Assume that $A$ satisfies $t$-NSP$(t, L, \bar{\gamma})$, where $t = |T|$ and $\bar{\gamma} < 1$. Let $x^*$ be the solution of (4) and $\bar{x}$ be the true signal. Then, $\|x^*_T\|_1 - \|\bar{x}_T\|_1 \leq 0$, and*

$$\|x^* - \bar{x}\|_1 \leq 2C_T \cdot \sigma_L(\bar{x}_T)_1, \tag{20}$$

*where*

$$C_T = \frac{1 + (1 + \max\{1, |T^C|/L\})\bar{\gamma}}{1 - \bar{\gamma}}.$$

*Proof.* For notation cleanliness, we introduce

$$S_1 := T^C = \mathcal{I},\ \ S_2 := S \subset T,\ \ S_3 := T \cap S^C,$$

which form a partition of $\{1, \ldots, n\}$.

Case 1: $|S_1| \leq L$. We can find $S' \subset S_2$ such that $|S_1 \cup S'| = L$. From $t$-NSP$(t, L, \bar{\gamma})$ for $A$, we get

$$\|(z-z')_{S_1}\|_1 \leq \|(z-z')_{S_1\cup S'}\|_1 \leq \bar{\gamma}\|(z-z')_{S_3}\|_1. \tag{21}$$

Case 2: $|S_1| > L$. Let $S'' \subset S_1$ denote the set of indices corresponding to the largest $L$ entries of $(z-z')_{S_1}$. From $t$-NSP$(t, L, \bar{\gamma})$ for $A$,, we have

$$\|(z-z')_{S_1}\|_1 \leq \frac{|S_1|}{L}\|(z-z')_{S''}\|_1 \leq \frac{|S_1| \cdot \bar{\gamma}}{L}\|(z-z')_{S_3}\|_1 \tag{22}$$

Combining (21) and (22) gives

$$\|(z-z')_{S_1}\|_1 \leq \max\left\{1, \frac{|S_1|}{L}\right\}\bar{\gamma}\|(z-z')_{S_3}\|_1.$$

This, together with (18) and (19), gives

$$\begin{aligned}
\|z - z'\|_1 &= \|(z-z')_{S_1}\|_1 + \|(z-z')_{S_2}\|_1 + \|(z-z')_{S_3}\|_1 && (23)\\
&\leq \left(1 + (1 + \max\{1, |S_1|/L\})\,\bar{\gamma}\right)\|(z-z')_{S_3}\|_1 && (24)\\
&\leq C_T\left(\|z'_T\|_1 - \|z_T\|_1 + 2\sigma_L(z_T)_1\right). && (25)
\end{aligned}$$

Finally, let $z$ and $z'$ denote the true signal $\bar{x}$ and the solution $x^*$ of (4), respectively. The optimality of $x^*$ gives $\|x^*_T\|_1 - \|\bar{x}_T\|_1 \leq 0$, from which (20) follows. □

Theorem 3.3, states that the reconstruction error of (4) is bounded by the best $L$-term approximation error of $\bar{x}_T$ up to a multiple depending on $\bar{\gamma}$ and $|T^C|$. When $t = n$, it reduces to the existing result for BP established in [9]:

$$\|x^* - \bar{x}\|_1 \leq 2\frac{1+\gamma}{1-\gamma} \cdot \sigma_{L'}(\bar{x})_1, \tag{26}$$

when $A$ satisfies the NSP of order $L'$ for $\gamma \in (0,1)$.

To compare truncated BP with BP on their error bounds given in (20) and (26), respectively, we need to study the tail of $\bar{x}$. We claim that making correct detections alone is not sufficient to make (20) better than (26). To see this, assume that $\bar{\gamma} = \gamma$ in both bounds. Support detection makes $t = |T| < n$ and $|T^C| > 0$. As discussed at the end of Subsection 3.1 above, we get $L' \leq L$. Then, it is unclear whether we have $\sigma_L(\bar{x}_T)_1 < \sigma_{L'}(\bar{x})_1$ or the other way. In addition, $C_T$ is bigger than $(1+\gamma)/(1-\gamma)$. Hence, the error bound in (20) can be bigger than (26). Only if the tail decays fast enough in the sense that $\sigma_L(\bar{x}_T)_1 \ll \sigma_{L'}(\bar{x})_1$, can (20) reliably give a smaller bound than (26). This comparison also applies to two instances of truncated BP, one having a bigger $T$ than the other. For support detection to lead to a reduced reconstruction error, there must be enough correct detections *and* $\bar{x}$ must have a fast decaying distribution of nonzeros. This conclusion matches our numerical results given in Section 5 below.

### 3.4 Iterative Behavior of ISD

The results in the above two subsections concern signal reconstruction by (4) not the ISD iterations, and exact reconstruction requires the $t$-NSP with a parameter $\bar{\gamma} < 1$. This subsection presents a sufficient condition for the ISD iterations to yield a decreasing sequence of $\bar{\gamma}$.

**Theorem 3.4.** *Suppose that $A$ has the t-NSP$(t, L, \bar{\gamma})$ as well as t-NSP$(t', L', \bar{\gamma}')$ with $t' < t$ and $L' < L$. If $(L - L') > \bar{\gamma}(t - t' - (L - L'))$, then $\bar{\gamma}' < \bar{\gamma}$.*

*Proof.* Let $0 < J' < J$ and $1 < \gamma < \infty$. For given $T'$, $\eta'$, and $S'$ that satisfy $T' \subset \{1, \ldots, n\}$, $|T'| = t'$, $\eta' \in \mathcal{N}(A)$, $\eta' \neq \mathbf{0}$, $S' \subset T'$, $|S'| = J'$, $\bar{\gamma}' = \|\eta'_{S'}\|_1 / \|\eta'_{T'\setminus S'}\|_1$, we have

$$\begin{aligned}
\|\eta'_{T'\setminus S'}\|_1 &= \|\eta'_{T\setminus S}\|_1 - \|\eta'_{T\setminus S-(T'\setminus S')}\|_1 \\
&\geq \bar{\gamma}^{-1}\|\eta'_S\|_1 - \|\eta'_{T\setminus S-(T\setminus S')}\|_1 \\
&= \bar{\gamma}^{-1}\|\eta'_{S'}\|_1 + \bar{\gamma}^{-1}\|\eta'_{S-S'}\|_1 - \|\eta'_{T\setminus S-(T\setminus S')}\|_1,
\end{aligned}$$

for any $S$ satisfying $S \supseteq S'$, $|S| = J$, and $S \subset T$, $|T| = t$, $S - S' \subseteq T \setminus T'$. In particular, we choose $S$ such that $S - S'$ consists of the largest $J - J'$ entries of $\eta'_{T\setminus T'}$ in magnitude.

According to $(J - J') > \bar{\gamma}(t - t' - (J - J'))$, we have

$$|S - S'| > \bar{\gamma}|T \setminus S - (T' \setminus S')|. \tag{27}$$

If $\eta'_{S-S'} \neq \mathbf{0}$, then this condition means $\bar{\gamma}^{-1}\|\eta'_{S-S'}\|_1 > \|\eta'_{T\setminus S-(T'\setminus S')}\|_1$ and, thus, $\|\eta'_{T'\setminus S'}\|_1 > \bar{\gamma}^{-1}\|\eta'_{S'}\|_1$. Otherwise, i.e., $\eta'_{S-S'} = \mathbf{0}$, then we have $\bar{\gamma}^{-1}\|\eta'_{S-S'}\|_1 = \|\eta'_{T\setminus S-(T'\setminus S')}\|_1 = 0$. However, we can still show $\|\eta'_{T'\setminus S'}\|_1 > \bar{\gamma}^{-1}\|\eta'_{S'}\|_1$ by showing $\|\eta'_{T\setminus S}\|_1 > \bar{\gamma}^{-1}\|\eta'_S\|_1$, i.e., the first inequality in the equation array above holds strictly. To see this, we first get $\eta'_{T'\setminus S'} \neq \mathbf{0}$ from $\bar{\gamma}^{-1}\|\eta'_{S-S'}\|_1 = \|\eta'_{T\setminus S-(T'\setminus S')}\|_1 = 0$, $\|\eta'_{T'\setminus S'}\|_1 \geq \bar{\gamma}^{-1}\|\eta'_{S'}\|_1$, $\bar{\gamma}^{-1} > 0$, and $\eta' \neq \mathbf{0}$. Next, we generate $\bar{S}$ by first letting it be $S$, second dropping any one entry in $S - S'$ (which has a zero value), and picking up a nonzero entry in $T' \setminus S'$. Such $\bar{S}$ satisfies

$$\|\eta'_{T\setminus S}\|_1 > \|\eta'_{T\setminus \bar{S}}\|_1 \geq \bar{\gamma}^{-1}\|\eta'_{\bar{S}}\|_1 > \bar{\gamma}^{-1}\|\eta'_S\|_1.$$

Therefore, we have

$$(27) \Rightarrow \|\eta'_{T'\setminus S'}\|_1 > \bar{\gamma}^{-1}\|\eta'_{S'}\|_1$$

Therefore $\bar{\gamma}' < \bar{\gamma}$. □

To understand the result, we assume that BP (ISD iteration 0) fails to reconstruct $\bar{x}$ so that we can apply $(t, L, \bar{\gamma})$ and $(t', L', \bar{\gamma}')$ to ISD iterations 0 and 1, respectively. Recall that the numbers of correct and wrong detections are denoted by $d_c$ and $d_w$, respectively. We can see $(L - L') > \bar{\gamma}(t - t' - (L - L'))$ being equivalent to $d_c > \bar{\gamma} d_w$, meaning that from $x^{(0)}$ ISD must make correct detections at least $\bar{\gamma}$ times as many as wrong detections in order to guarantee $\bar{\gamma}' < \bar{\gamma}$, which indicates a forward step toward exact reconstruction. The result can also be applied to two consecutive ISD iterations $s$ and $s + 1$ and give the condition $\Delta d_c > \bar{\gamma}\Delta d_w$, where $\bar{\gamma}$ applies to iteration $s$ and $\Delta d_c$ and $\Delta d_w$ are the changes of correct and wrong detections in number from iteration $s$ to $s + 1$, respectively.

In practice, ISD does not know $\bar{\gamma}$, $d_c$, or $d_w$, but ISD needs to know when to stop its iterations. We suggest two alternatives. The first one uses the $(\lfloor m/2 \rfloor + 1)'$th largest component of $x^{(s)}$. According to [1], this value is proportional to an upper bound of the error. Therefore, once the value is 0 or small enough, we obtain an (almost) exact reconstruction and can stop the ISD iteration. In addition, we can also stop the ISD iteration if this value stagnates or shows a steady increase over the past few iterations. Another stopping rule is based on comparing $I^{(s-1)}$ and $I^{(s)}$. When ISD fails to further improve the solution (including the case the solution is already exact), $I^{(s)}$ and $I^{(s-1)}$ will be (almost) identical.

# 4 Threshold–ISD for Fast Decaying Signals

## 4.1 A Support Detection Scheme Based on Thresholding

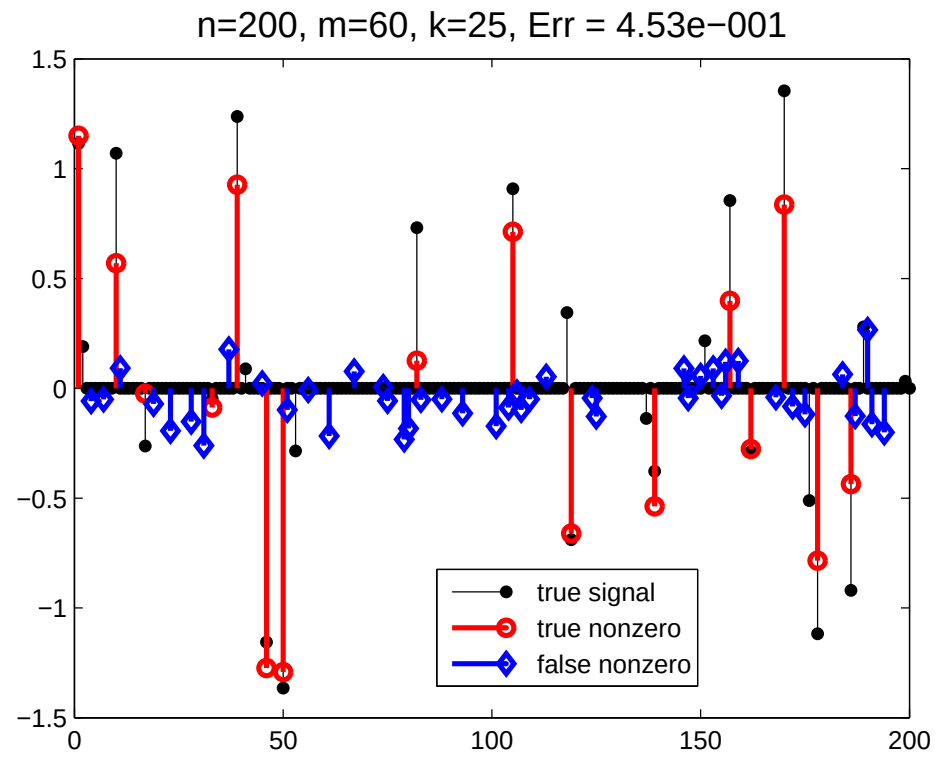

(a) The nonzeros of the true signal v.s. the true/false nonzeros of the reconstructed signal

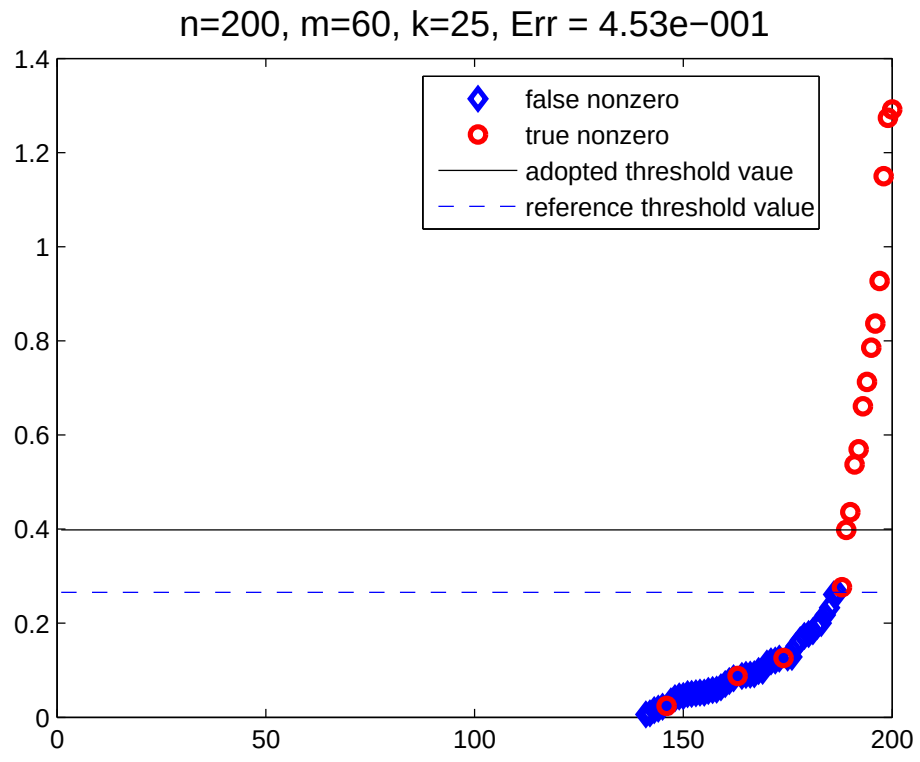

(b) The sorted components of the failed reconstruction and their first significant jump

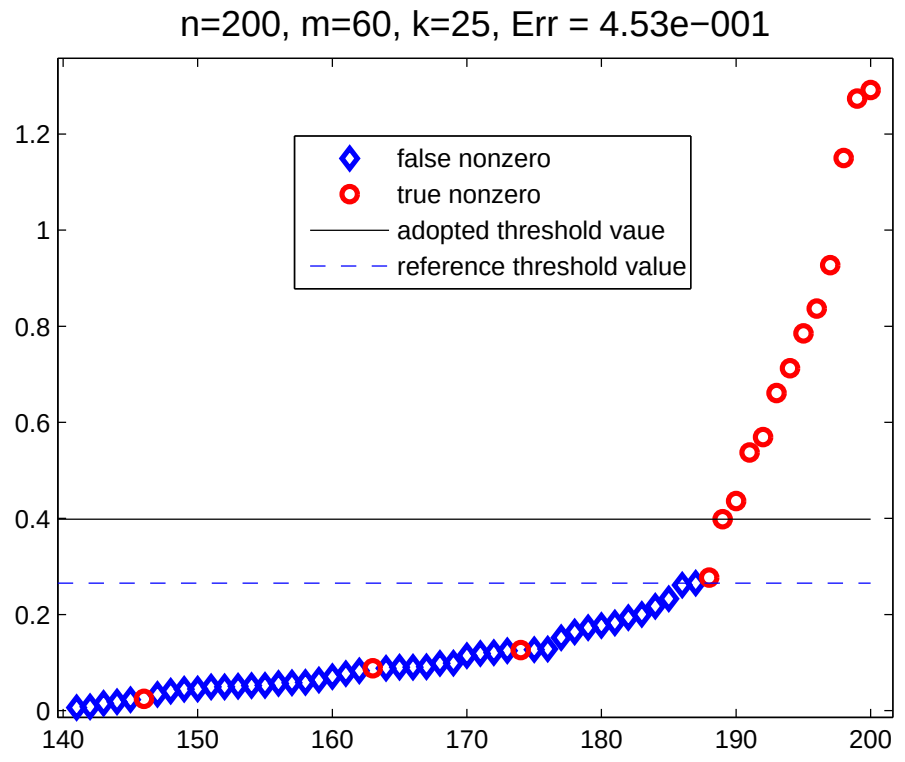

(c) A zoom-in of Figure (b)

Figure 2: Illustration of support detection from a failed reconstruction by looking for the "first significant jump". The failed reconstruction was obtained by BP from Gaussian linear measurements of a sparse Gaussian signal. "Err" is the relative error in $\ell_2$ norm.

ISD requires reliable support detection. In this section, we present effective detection strategies for signals with a fast decaying distribution of nonzero values (hereafter, we call them *fast decaying signals*), which include sparse Gaussian signals and certain power–law decaying signals. Our strategies are based on thresholding

$$I^{(s+1)} := \{i : |x_i^{(s)}| > \epsilon^{(s)}\}, \quad s = 0, 1, 2, \dots. \tag{28}$$

We term the resulting algorithm *threshold–ISD*. Before discussing the choice of $\epsilon^{(k)}$, we note that the support sets $I^{(s)}$ are not necessarily increasing and nested, i.e., $I^{(s)} \subset I^{(s+1)}$ may not hold for all $s$. This is important because it is very difficult to completely avoid wrong detections by setting $\epsilon^{(s)}$ based on available intermediate solutions $x^{(i)}$, for $i \leq s$. The true solution is not known, and a component of $x^{(i)}$, no matter how big, could nevertheless still be zero in the true solution. Not requiring $I^{(s)}$ to be monotonic leaves the chance for support detection to remove previous wrong detections, making it less sensitive to $\epsilon^{(s)}$ and thus making $\epsilon^{(s)}$ easier to choose.

We study different rules for $\epsilon^{(s)}$. We have seen in Section 2 the simple rule $\epsilon^{(s)} = \|x^{(s)}\|_\infty/\beta^{(s+1)}$ with $\beta > 0$. This rule is quite effective with an appropriate $\beta$. However, the proper range of $\beta$ is case-dependent. An excessively large $\beta$ results in too many false detections and consequently low solution quality whilst an excessively small $\beta$ tends to cause a large number of iterations. Because the third rule below is more effective, this first rule is not recommended.

The second rule is a toll–based rule, namely, setting $\epsilon^{(s)}$ so that $I^{(s)}$ has a given cardinality, which increases in $s$. We tried cardinality sequences such as $1, 2, \ldots$ and $2, 4, \ldots$ and obtained high quality reconstructions from a small number of measurements. However, because $I^{(s)}$ grows slowly, threshold–ISD takes a large number of iterations. In comparison, the third rule below offers a much better balance between quality and speed.

The rule of our choice is based on locating the "first significant jump" in the increasingly sequence $|x^{(s)}_{[i]}|$ ($x_{[i]}$ denotes the $i$th largest component of $x$ by magnitude), as illustrated by Figure 2. The rule looks for the smallest $i$ such that

$$|x^{(s)}_{[i+1]}| - |x^{(s)}_{[i]}| > \tau^{(s)}, \tag{29}$$

where $\tau^{(s)}$ is selected below. This amounts to sweeping the increasing sequence $|x^{(s+1)}_{[i]}|$ and looking for the first jump larger than $\tau^{(s)}$. In the example in Figure 2, it is located at $[i] = 188$. Then, we set $\epsilon^{(s)} = |x^{(s)}_{[i]}|$. Typically, this rule allows (28) to detect a large number of true nonzeros with few false alarms. It is less arbitrary than the first rule while also leading to faster convergence than the second rule.

The "first significant jump" exists because in $x^{(s)}$, the true nonzeros (the red circles in Figure 2) are large in size and small in number, while the false ones (the blue diamonds in Figure 2) are large in number and small in size. Therefore, the magnitudes of the true ones are spread-out while those of the false ones are clustered. The false ones are the smearing due to the nonzeros in $\bar{x}$ that are zeroed out in $x^{(s)}$. To explain this mathematically, let us decompose $x^{(s)}$, which we assume to have $m$ nonzero entries[2]. Slightly abusing the notation, we let $B = \{i : x^{(s)}_i \neq 0\}$ denote both the set of basic indices and the corresponding square submatrix of $A$ so we have $Bx^{(s)}_B = b$. Let $V = B^C$. From basic algebra, we get $x^{(s)}_B = \bar{x}_B + B^{-1}A_V\bar{x}_V$ where $\bar{x}_B = [\bar{x}_U; \mathbf{0}]$ for $U = B \cap \text{supp}(\bar{x})$. Because $\bar{x}_V$ tends to have smaller components than $\bar{x}_U$[3] and $B^{-1}$ has a diluting effect, the components of $B^{-1}A_V\bar{x}_V$ tend to be much smaller than those in $\bar{x}_U$. Therefore, $x^{(s)}_U$ are typically dominated by $\bar{x}_U$ and thus have relatively large components.

Here we adopt a very simple method to define $\tau^{(s)}$ for different kinds of sparse or compressible signals. For the sparse Gaussian signals, we set $\tau^{(s)}$ simply as $m^{-1}\|x^{(s)}\|_\infty$, which is used in the example shown in Figure 2 and gives no false detection. For the first 5 iterations in the experiments reported in Section 5 below, we made the detection slightly more conservative by increasing $\tau^{(s)}$ to $\frac{6}{s+1}\tau^{(s)}, s = 0, 1, 2, 3, 4$.

Besides sparse Gaussian signals, we also tried the "first significant jump" rule on synthetic sparse and compressible power–law decaying signals, as well as on the wavelet coefficients of the Shepp-Logan phantom and the cameraman image in Section 5. The sparse and compressible power–law decaying signals were constructed by first generating a sequence of numbers obeying the power-decay rule like $\{i^{-1/\lambda}\}_{i=1}^k$ followed by multiplying each entry by a random sign and applying a random permutation to the signed sequence. We set $\tau^{(s)}$ according to $\lambda$. For example, when $\lambda = 1/3$, one can set $\tau^{(s)} = \|x^{(s)}\|_\infty/m/20$; when $\lambda = 0.8$, or 1, one can set $\tau^{(s)} = \|x^{(s)}\|_\infty/m/5$; when $\lambda = 2$ or 4, one can set $\tau^{(s)} = \|x^{(s)}\|_\infty/m/2$. For the wavelet coefficients of the phantom image and cameraman image, one can set $\tau^{(s)} = \|x^{(s)}\|_\infty/m$. Like we did for

[2] (4) can be reduced to a linear program. It is possible that $x^{(s)}$ has either less or more than $m$ nonzero entries but for most matrices $A$ used in CS and most signals, $x^{(s)}$ has exactly $m$ nonzero entries with a nonsingular basis $B$. This assumption offers technical convenience and is not essential to the argument that follows.

[3] Roughly, a fast-decaying $\bar{x}$ can be regarded to consist of a signal part of large components and a noise part of smaller and zero components. Stability results say that the signal part are mostly included in $\bar{x}_U$.

sparse Gaussian signals, for the first 9 iterations in the experiments, we made the detection slightly more conservative by increasing $\tau^{(s)}$ to $\frac{8}{s+1}\tau^{(s)}$, $s = 0, 1, 2, 3, 4, 5, 6, 7$. The above heuristics, which worked very well in our experiments, is certainly not necessarily optimal; on the other hand, it has been observed that threshold–ISD is not very sensitive to $\tau^{(s)}$.

Finally, we give three comments on support detections. First, one can apply other available effective jump detection methods [28, 21]. Second, any threshold-based support detection rule requires true signals to have a fast decaying distribution of nonzeros in order to work reliably. It does not work on signals that decay slowly or have no decay at all (e.g., sparse Bernoulli signals). Third, one should be able to find better support detection methods for real world signals such as those with grouped nonzero components (cf. the model-based CS [2]) and natural/medical images (cf. edge-enhanced compressive imaging [19]).

### 4.2 YALL1 and Warm-Start

The truncated BP model (4) can be solved by most existing $\ell_1$ algorithms/solvers with straightforward modifications. We chose YALL1 [30] since it is among the fastest ones whether the solution is sparse or not. In the numerical tests reported in Section 5 below, we applied YALL1 to all BP, truncated BP, and weighted $\ell_1$ problems that arise in threshold–ISD and the compared algorithms.

For completeness, we give a short overview of YALL1. Based upon applying the alternating direction method to the Lagrange dual of

$$\min_x \{\sum_{i=1}^n w_i |x_i| : Ax = b\},$$

the iteration in YALL1 has the basic form:

$$y^{l+1} = \alpha A z^l - \beta(Ax^l - b), \tag{30}$$

$$z^{l+1} = P_w(A^* y^{l+1} + x^l/\mu), \tag{31}$$

$$x^{l+1} = x^l + \gamma\mu(A^* y^{l+1} - z^{l+1}), \tag{32}$$

where $\mu > 0$, $\gamma \in (0, (1+\sqrt{5})/2)$, $\alpha = 1$ and $\beta = \frac{1}{\mu}$. $P_w$ is an orthogonal projection onto the box $\mathbf{B}_w \triangleq \{z \in \mathbb{C}^n : |z_i| \le w_i, i = 1, \ldots, n\}$. This is a first-order primal–dual algorithm in which the primal variables $x$ and dual variables $y$ and $z$ (a dual slack) are updated at every iteration. In our numerical experiments, we stopped YALL1 iterations once the relative change $\|x^{l+1} - x^l\|_2 / \|x^l\|_2$ fell below a certain prescribed stopping tolerance.

YALL1 was called multiple times in threshold–ISD to return $x^{(s)}$, $s = 0, 1, \ldots$. Because the "first significant jump" rule does not need a highly accurate solution, a loose stopping tolerance was set in YALL1 for all but the last threshold–ISD iteration. The tolerances used are given in Subsection 5.5 below. To determine whether a threshold–ISD iteration $k$ was the final iteration, the loose stopping tolerance was put in place to stop YALL1 and upon stopping of YALL1, $I^{(s+1)}$ was generated and compared to $I^{(s)}$ and $I^{(s-1)}$. If all three $I$'s were identical or almost so, any further iteration would likely give the same or a very similar solution. At this time, we let YALL1 resume from where it stopped and return an accurate final solution till a tighter tolerance was reached. In other words, threshold–ISD uses coarse solutions for support detection and returns a fine solution to the user.

To further accelerate threshold–ISD, we warm-started YALL1 whenever possible. Specifically, for each instance of BP, truncated BP, and weighted $\ell_1$ problems, YALL1 was started from the solution $(x, y, z)$ of the previous instance if available.

As a result of applying varying stopping tolerance and warm–start, the total threshold–ISD time, including the time of solving multiple truncated BP problems, was almost the same on average as the time of solving a single BP problem.

## 5 Numerical Implementation and Experiments

Threshold–ISD was compared to the BP model (2), the iterative reweighted least-squares algorithm (IRLS) [8], and the iterative reweighted $\ell_1$ minimization algorithm (IRL1) [7]. IRLS and IRL1 appear to be state-of-the-art in terms of the number of measurements required. These comparisons[4] show that threshold–ISD requires as few measurements as IRLS and meanwhile runs as fast as BP, which is much faster than IRLS.

## 5.1 Reviews of IRL1 and IRLS

Let us first briefly review the algorithms IRL1 [7] and IRLS [8]. They are both iterative procedures attempting to solve the following $\ell_p$ minimization problem:

$$\min \|x\|_p \quad \text{s.t.} \quad Ax = b \tag{33}$$

where $p \in [0, 1]$. At the $s$-th iteration, the IRL1 algorithm computes

$$x^{(s)} \leftarrow \min_x \{\sum_{i=1}^{n} w_i^{(s)} |x_i| : Ax = b\}, \tag{34}$$

where the weights are set as

$$w_i^{(s)} := (|x_i^{(s-1)} + \eta|)^{p-1}, \tag{35}$$

and $\eta$ is a regularization parameter. Initially, $x^{(0)}$ is the solution of the BP problem.

IRLS iteratively minimizes a weighted $\ell_2$ function to generate $x^{(s)}$:

$$x^{(s)} \leftarrow \min_x \{\sum_i \tilde{w}_i^{(s)} |x_i|^2 : Ax = b\}, \tag{36}$$

The solution of (36) can be given explicitly as

$$x^{(s)} = Q_s A^{\mathrm{T}} (A Q_s A^{\mathrm{T}})^{-1} b \tag{37}$$

where $Q_s$ is the diagonal matrix with entries $1/\tilde{w}_i^{(s)}$ and the weights are set as

$$\tilde{w}_i^{(s)} := (|x_i^{(s-1)}|^2 + \zeta)^{p/2-1} \tag{38}$$

and $\zeta$ is a regularization parameter. Initially, $x^{(0)}$ is the least-squares solution of $Ax = b$.

We set $p := 0$ uniformly in the experiments since it is reported that this value leads to better reconstructions than $p > 0$. Notice that if $x^{(s-1)}$ in both (35) and (38) are set equal to $x$ and $\eta = \zeta = 0$, then following the convention $0/0 = 0$, we have $\sum_i w_i^{(s)} |x_i|^2 = \sum_i \tilde{w}_i^{(s)} |x_i|^2 = \|\bar{x}\|_0$. This result, though not holding for $\eta, \zeta > 0$, indicates that the two objective functions are smooth approximations to $\|x\|_0$. When $\eta$ and $\zeta$ are large, $\sum_{i=1}^n w_i^{(s)} |x_i|$ and $\sum_i \tilde{w}_i^{(s)} |x_i|^2$ are close to $\|x\|_1$ and $\|x\|_2^2$, respectively, so they tend to have fewer local minima. Therefore, $\eta$ and $\zeta$ were both initially large and gradually reduced as $s$ increased. The setting of these parameters were given in Subsection 5.4 below.

## 5.2 Denoising Minimization Problems

The measured data is sometimes inaccurate due to various kinds of imprecisions or contaminations. Assume that $b = Ax + z$, where $z$ is i.i.d. Gaussian with zero mean and standard deviation (noise level) $\sigma$. When $\sigma$ is not big, the unconstrained BP problem (3) is known to yield a faithful reconstruction for an appropriate $\rho$ depending on $\sigma$. We found that (3) could be solved faster and yield a slightly more accurate solution than the constrained BP problem (2). Therefore, (3) was used in our tests with noisy measurements. In our figures, we use "L1/L2" for (3).

For IRL1, reweighting was applied to the above noise–aware problem (3), and each of its iteration was changed from (34) to

$$x^{(s)} \leftarrow \min_x \{\sum_{i=1}^{n} w_i^{(s)} |x_i| + \frac{1}{2\rho} \|b - Ax\|_2^2\}, \tag{39}$$

[4] More comparisons to other algorithms include certain greedy algorithms are given on the second author's website.

where weights $w_i^{(s)}$ were generated as before. Given an index set $T$, threshold–ISD solved the following truncated $\ell_1$ version of (3):

$$x^{(s)} \leftarrow \min \|x_{T^{(s)}}\|_1 + \frac{1}{2\rho}\|b - Ax\|_2^2. \tag{40}$$

The same $\rho$ was set for the three problems (3), (39) and (40), which were all solved by YALL1 iterations (30), (31) and (32) with new $\alpha := \frac{\mu}{\mu+\rho}$ and $\beta := \frac{1}{\mu+\rho}$.

For IRLS, however, we did not relax $Ax = b$ in (36) for noisy measurements because the resulting unconstrained problem is no easier to solve and neither does it return solutions with less error, at least when the error level $\sigma$ is not excessively large. Therefore, (36) was solved by IRLS for noisy measurements.

## 5.3 Transform Sparsity

In many situations, it is not the true signal $\bar{x}$ itself but its representation under a certain basis, frame, or dictionary that is sparse or compressible. In such a case, $\bar{y} = W\bar{x}$ is sparse or compressible for a certain linear transform $W$. Fwavelets, curvelets, etc. Instead of minimizing $\|x\|_1$ and $\|x_T\|_1$, $\|Wx\|_1$ and $\|(Wx)_T\|_1$ should be minimized, respectively. Then, the weight variables in IRL1 and IRLS should be updated according to the components of $(Wx)$ instead of those of $x$. In case of transform sparsity, the above simple changes were applied to all algorithms and solvers.

## 5.4 Experimental Settings and Test Platforms

| # | Nonzeros or Image Name | Noise $\sigma$ | Dimension $n$ | Sparsity $k$ | Measurements $m$ | Repetitions |
|---|---|---|---|---|---|---|
| 1 | Gaussian | 0 | 600 | 8 | 16:4:100 | 100 |
|  | Gaussian | 0 | 600 | 40 | 80:10:220 | 100 |
|  | Gaussian | 0 | 600 | 150 | 250:10:400 | 100 |
| 2 | Gaussian | 0 | 3000 | 100 | 200:50:800 | 100 |
| 3 | Gaussian | 0.0001 | 2000 | 100 | 325 | 200 |
|  | Gaussian | 0.001 | 2000 | 100 | 325 | 200 |
|  | Gaussian | 0.01 | 2000 | 100 | 325 | 200 |
| 4 | Power–Law at varying rates | 0 | 600 | 40 or 600 | varied | 100 |
| 5 | Shepp–Logan Phantom | 0 | 128 × 128 | 1685 | 2359:337:6066 | 10 |
|  | Shepp–Logan Phantom | 0.001 | 128 × 128 | 1685 | 2359:337:7414 | 10 |
|  | Cameraman | 0 | 256 × 256 | 65536 | 6553:2621:32763 | 10 |

Table 1: Summary of test sets.

The test sets are summarized in Table 1. Our experiment included five different test sets, the first four of which used various synthetic signals and standard i.i.d. Gaussian sensing matrices $A$ generated by `A=randn(m,n)` in MATLAB. The first two sets used noise-free measurements and different amounts of white noise added to measurements in the third set. With noise, exact reconstruction was impossible, so in the third set we did not change $m$ but measured solution errors for different levels of noise. The fourth set used signals with entries following power laws among which a part of the signals had their tails truncated and thus sparse. As zero–decay sparse signals are just sparse $\pm 1$ signals, this set also included sparse Bernoulli signals. The last (fifth) set used two–dimensional images of different sizes and tested sensing matrices $A$ that were partial discrete cosine matrices formed by choosing the first and a random subset of the remaining rows from the full discrete cosine matrices. Since all the operations in threshold–ISD and IRL1 involving $A$ (which are $Ax$ and $A^\top x$) were computed by the discrete cosine transform, $A$ were never explicitly formed or stored in memory in these two algorithms. On the other hand, IRLS needs explicit matrices $A$ so it was not tested in the fifth set[5]. The fifth set also included tests with noise added to the measurements. The images were assumed to be sparse under the two–dimensional Haar wavelets.

Specifically, the sparse Gaussian signals were generated in MATLAB by

```
xbar =zeros(n,1);  p=randperm(n); xbar(p(1:k))=randn(k,1);
```

and the sparse Bernoulli signals were generated by the same commands except

[5]IRLS iterations could be modified to use $Ax$ and $A^\top x$ rather than $A$ in the explicit form, but for the purpose of this paper, no modification was done.

```
xbar(p(1:k))=2*(rand(k,1)>0.5)-1;
```

The power–law decaying signals were generated by

```
xbar=zeros(n,1); p=randperm(n);
xbar(p(1:n))=sign(randn(n,1)).*((1:n).^(-1/lambda))';
xbar=xbar/max(abs(xbar));
```

and replacing the second line by `xbar(p(1:k))=sign(randn(k,1)).*((1:k).^(-1/lambda))'` we obtained the sparse ones. Variable `lambda` was set to different values (described in Subsection 4.1 above), which controls the rate of decay. The larger `lambda`, the lower the rate of decay.

All test code was written and tested in MATLAB v7.7.0 running in GNU/Linux Release 2.6.9–55.0.2 on a Dell Optiplex GX620 with dual Intel Pentium D CPUs 3.20GHz (only one CPU was used by MATLAB) and 3 GB of memory.

## 5.5 Stopping Tolerance and Smoothing Parameters

Performances of all tested code depend on parameters. For fairness, threshold–ISD, IRL1, IRLS, and BP were stopped upon

$$\frac{\|x^{l+1} - x^l\|_2}{\|x^l\|_2} \leq \epsilon, \tag{41}$$

with different intermediate but the same final stopping tolerances $\epsilon$.

**Test sets 1, 2, and 4**: In all tests, threshold–ISD was set to run no more than 9 iterations (with the reason given below in test sets 1 and 2), in which the first iteration had $\epsilon := 10^{-1}$ and the rest except for the last one had $\epsilon := 10^{-2}$. Smaller $\epsilon$ values did not make solutions or running times better. $\epsilon := 10^{-6}$ was set for the final iterations of threshold–ISD. The stopping rule is based on comparing consecutive support sets $I^{(s)}$ as described in Section 4.2.

$\epsilon := 10^{-6}$ was also set for BP, all iterations of IRL1. Larger intermediate $\epsilon$ values would make IRL1 return worse solutions. IRL1 had 9 iterations as recommended in [7], but its smoothing parameter $\eta$ was initialized to 1 and reduced by half each time, different from but slightly better than the recommendation.

Recommended in [8] for IRLS and for test sets 1 and 2, its smoothing parameter $\zeta$ was initialized to 1 and reduced to $\zeta/10$ whenever (41) was satisfied for $\epsilon := \sqrt{\zeta}/100$ until $\zeta$ reaches $10^{-8}$ when the final $\epsilon = 10^{-6}$ became effective. We optimized $\epsilon$ and the stopping value of $\zeta$ for the more challenging test 4. For power–law decaying signals (either sparse or not), $\epsilon := 10^{-3/2}\sqrt{\zeta}$ and the stopping $\zeta$ was set to $10^{-9}$; for sparse Bernoulli signals, $\epsilon := 10^{-1}\sqrt{\zeta}$ and the stopping $\zeta$ was set to $10^{-10}$. Again, $\epsilon$ reached $10^{-6}$ finally. The above optimization to IRLS made its frequency of successful reconstruction slightly higher than threshold–ISD in a couple of tests.

**Test set 3**: Because of measurement noise and thus reconstruction errors, it was not necessary to impose a tight tolerance for this set of tests. All the four algorithms had the reduced final $\epsilon := \sqrt{\sigma}/100$ uniformly. Threshold–ISD ran no more than 9 iterations with the same intermediate $\epsilon$ values as in test sets 1, 2, and 4. IRL1 ran 9 iterations with intermediate $\epsilon := \sqrt{\sigma}/100$ constantly. For all but the last IRLS iteration, $\epsilon := \max\{\sqrt{\sigma}/100, \sqrt{\zeta}/100\}$.

**Test set 5**: Threshold–ISD was only compared to BP and IRL1 because matrices $A$ were too large to form in IRLS. The only parameter change was the final $\epsilon := \max\{10^{-4}, \sigma/10\}$ for all the three tested algorithms.

## 5.6 Experimental Results

**Test set 1:** Sparse signals containing $k = 40, 8, 150$ nonzeros were used in this set, and corresponding results were plotted in Figures Figures 3, 4, and 5, respectively.

Figure 3 depicts the performance of the four tested algorithms. Figure 3(a) shows that threshold–ISD and IRLS achieved almost the same recoverability, which was significantly higher than that of IRL1 in terms of the number of required measurements. With no surprise, the recoverability of the BP method was the worst. Figure 3(b) shows that threshold–ISD was much faster than both IRL1 and IRLS, and was even comparable to BP To sum up, threshold–ISD was not only the fastest one but also the one that required the least number of measurements.

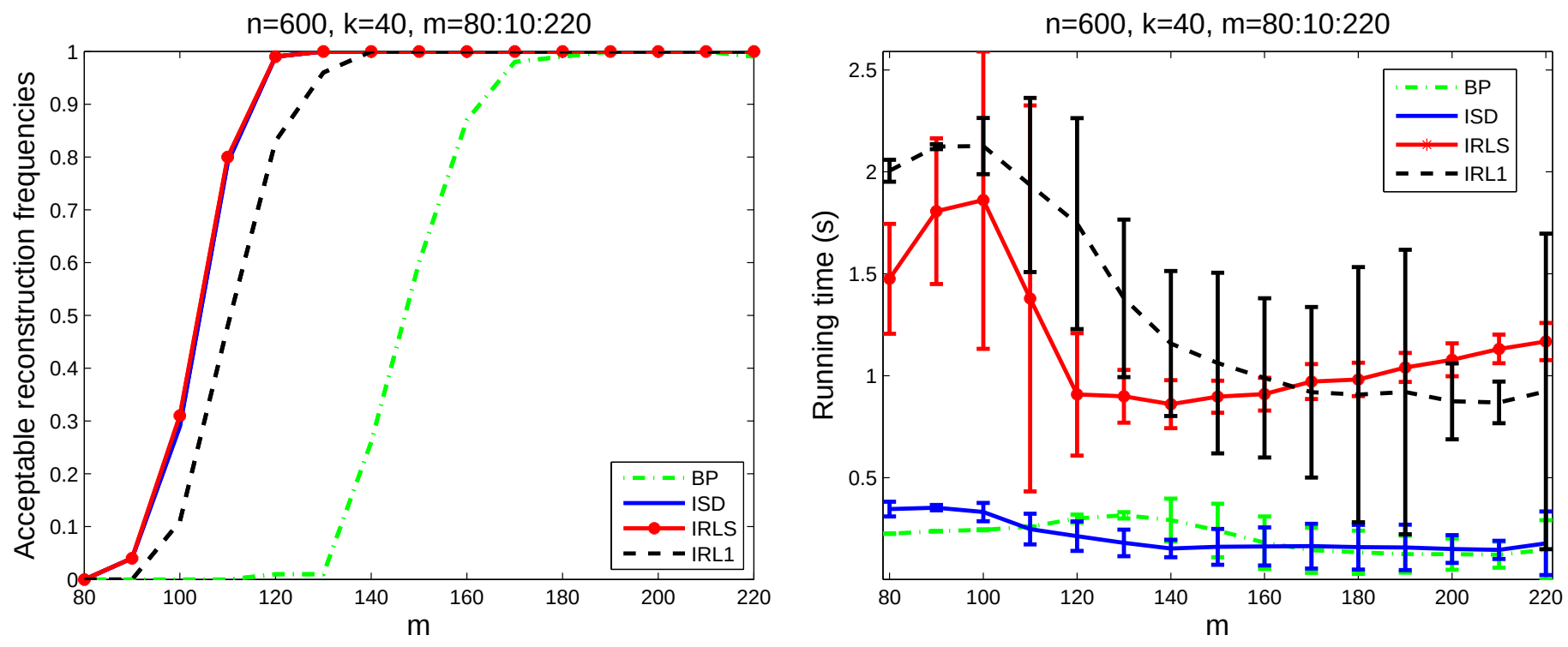

(a) Recoverability (b) CPU time

Figure 3: Test set 1 with $k = 40$: Comparisons in recoverability and CPU time

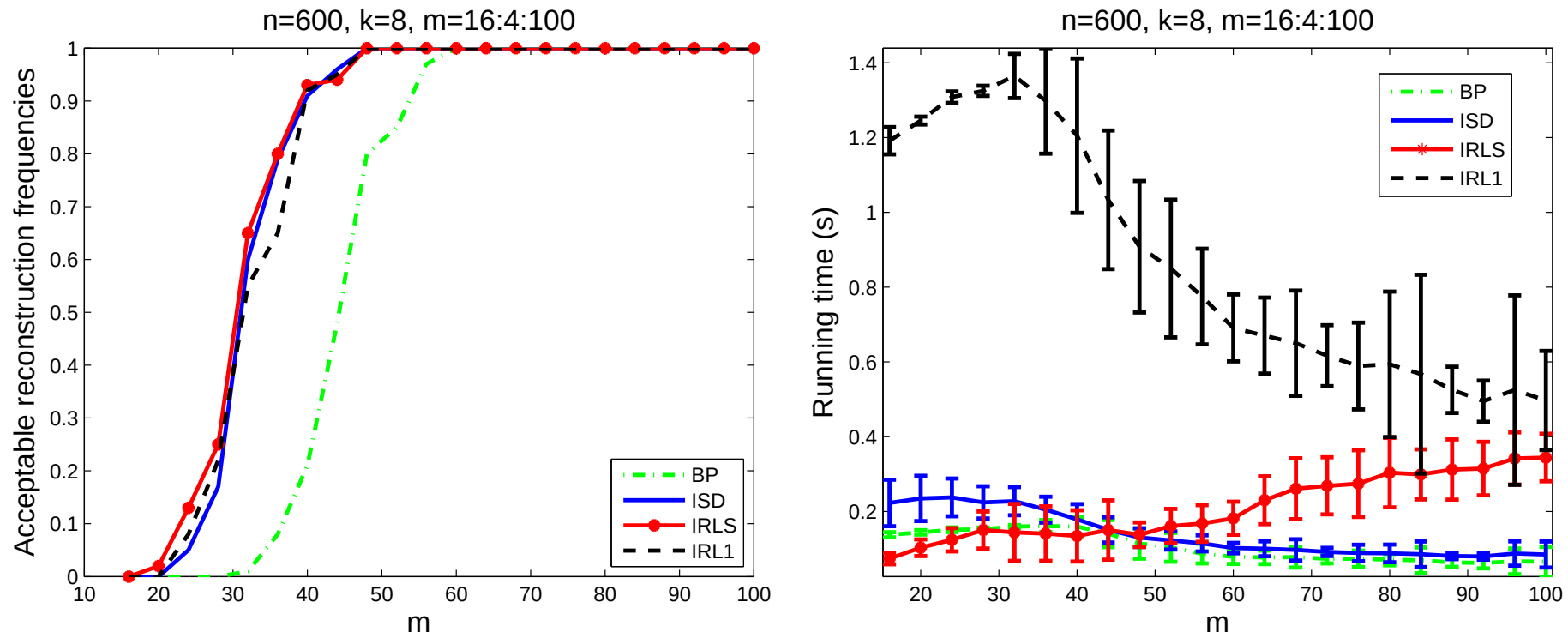

(a) Recoverability (b) CPU time

Figure 4: Test set 1 with $k = 8$: Comparisons in recoverability and CPU time

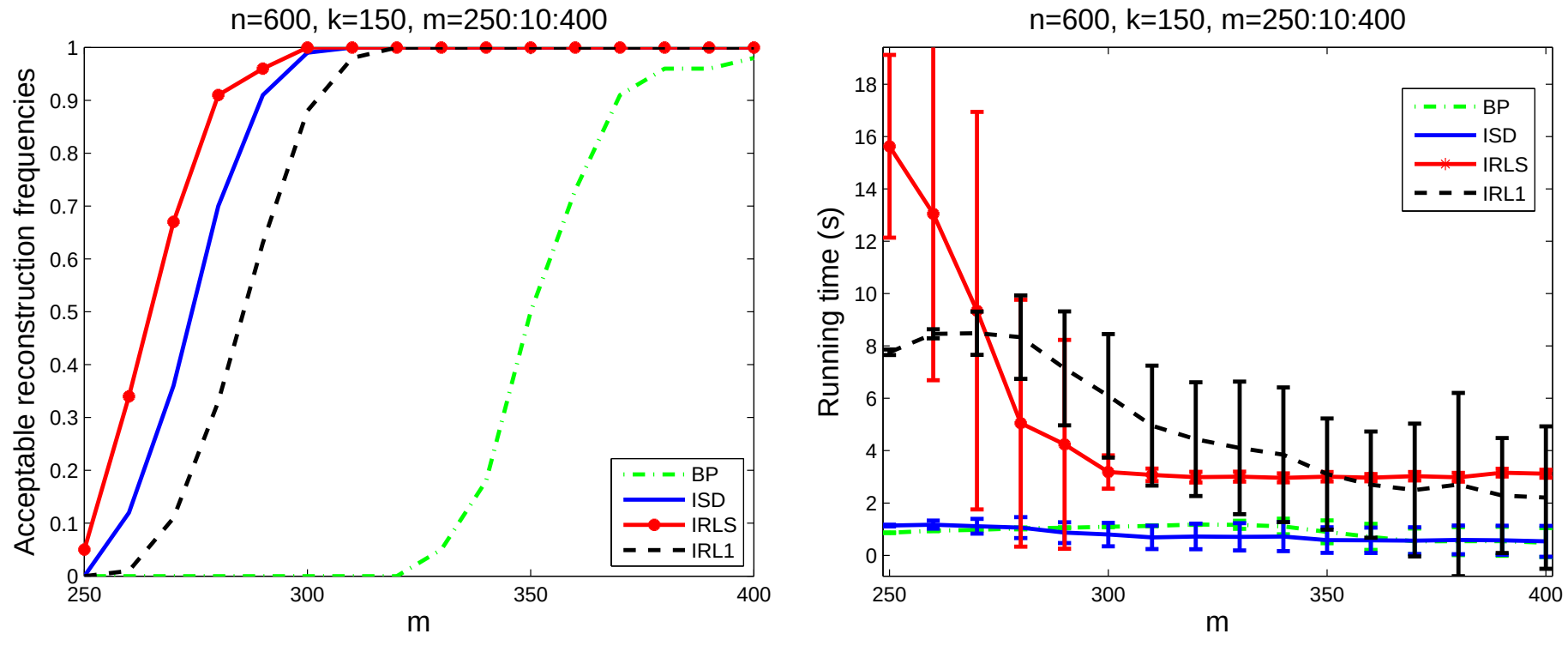

(a) Recoverability (b) CPU time

Figure 5: Test set 1 with $k = 150$: Comparisons in recoverability and CPU time

With the small $k = 8$ (Figure 4), threshold–ISD had no speed advantage over IRLS but it was still much faster than IRL1. Quality–wise, threshold–ISD was on par with IRLS and IRL1 and better than BP. With the larger $k = 150$, threshold–ISD was much faster than both IRLS and IRL1, and all three achieved comparable recoverability with 100% starting around $m = 300$.

**Test set 2:** This test set used larger signals ($n = 3000$). Figure 6 shows that threshold–ISD, IRLS, and IRL1 achieved similar recoverability as they did in test set 1. Because of the relatively large signal size, IRLS and IRL1 were however much slower than threshold–ISD. The fact that threshold–ISD ran as fast as BPsuggests that effective support detection accelerates subproblem solution (by YALL1) in threshold–ISD. The results also show that threshold–ISD was scalable to both signal and measurement sizes.

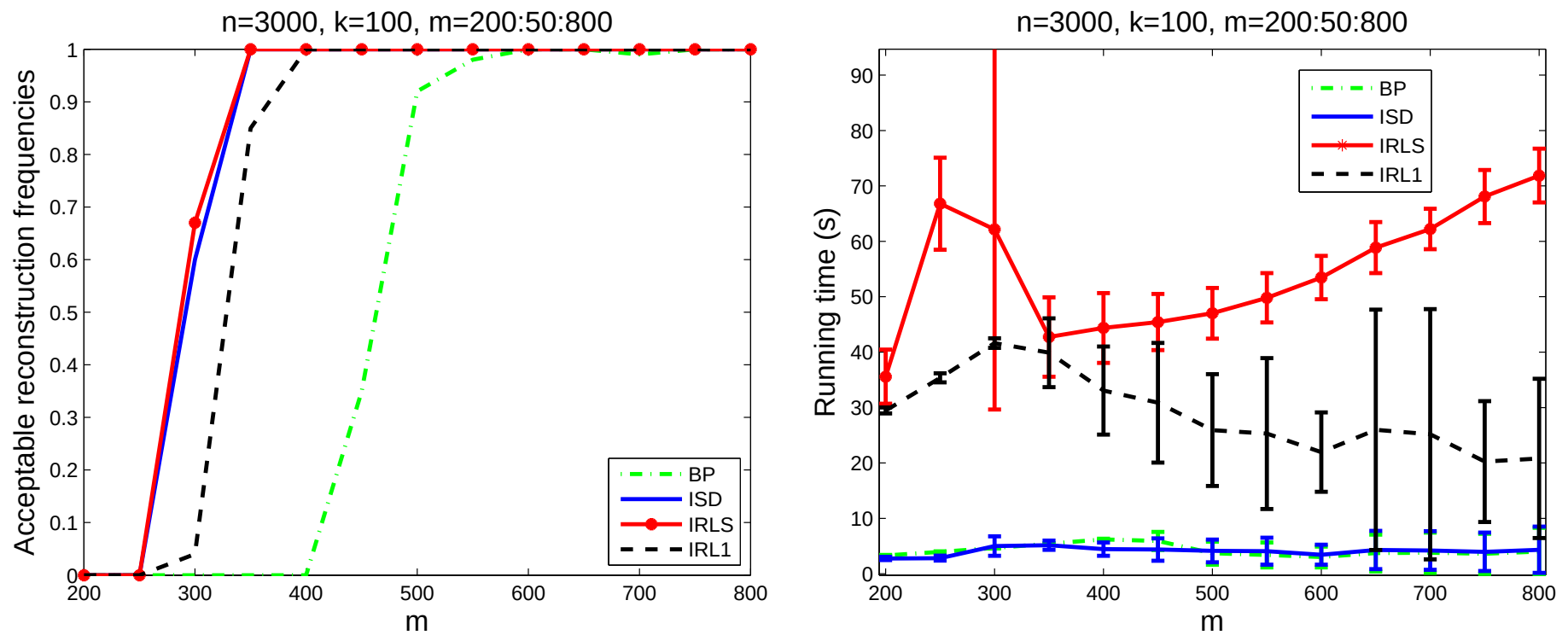

(a) Recoverability (b) CPU time

Figure 6: Test set 2: Comparisons in recoverability and CPU time

**Test set 3:** This test set compared solution times and errors of the tested algorithms given noisy measurements with three noise levels: $\sigma = 0.0001$, $\sigma = 0.001$ and $\sigma = 0.01$. The corresponding results are depicted in Figures 7, 8, and 9, respectively, each including three subplots for CPU times, $\ell_2$ and $\ell_1$ errors. It is clear from these figures that threshold–ISD was significantly faster than IRLS and IRL1 and slightly faster than BP. Threshold–ISD and IRLS were on par on solution quality except that at $\sigma = 0.01$, threshold–ISD had two fails among the total two hundred trials. IRL1 had much more fails, and BP was the worst. Taking both reconstruction errors and CPU times into consideration, threshold–ISD appeared to the best.

**Test set 4:** This test set included two subsets. The first subset (Figure 10) used sparse signals with nonzeros decaying in power–laws at rates $\lambda = 1, 2, 4$, as well as sparse Bernoulli signals. The second subset (Figure 11) used two compressible (non–sparse) signals with nonzeros decaying in power–laws at rates $\lambda = 1/3, 0.8$, as well as their sparse tail–removed versions obtained by removing all but the largest $k = 8, 40$ entries, respectively.

In the first subset of tests, threshold–ISD, IRL1, and IRLS had better recoverability than BP, but the advantage diminished as $\lambda$ increased (i.e., the rate of decay decreased). In cases of zero–decay where the signals were sparse Bernoulli, all the tested algorithms had similar recoverability. These results match the conclusion in our theoretical analysis that thresholding–based support detection is effective only if the nonzeros of signals have a fast–decaying distribution.

The second subset of tests show how the tail of a compressible signal affects its reconstruction. By comparing Figures 11 (c) with (d) and (e) with (f) (i.e., tail–free v.s. tailed signals), we can observe that it was much easier for any of the tested algorithms to reconstruct a tail–free sparse signal than a compressible signal. In addition, because the signal corresponding to $\lambda = 0.8$ has a larger tail, the improvement of threshold–ISD, IRL1, and IRLS over BP was quite small.

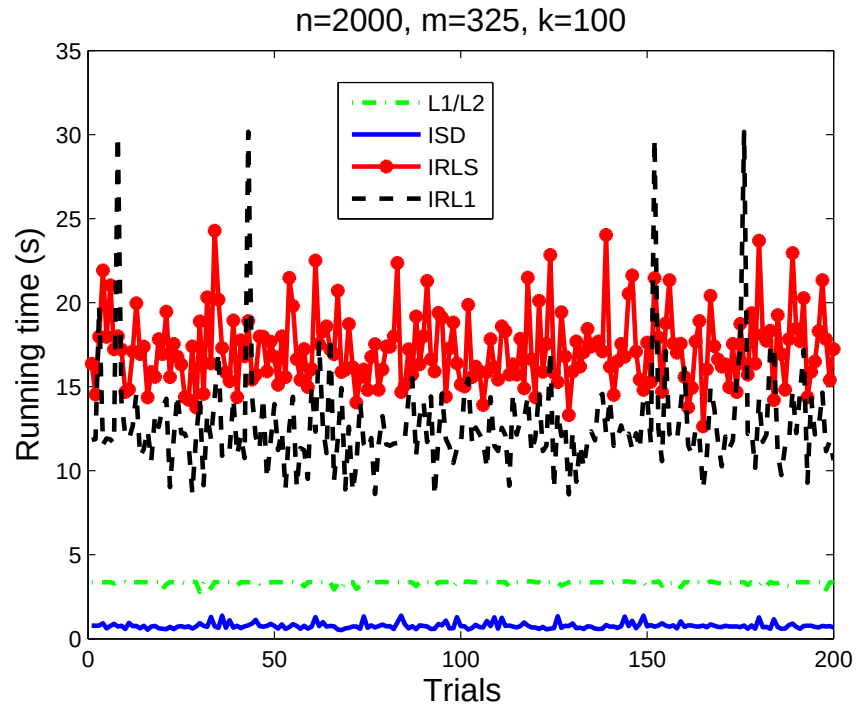

(a) CPU time

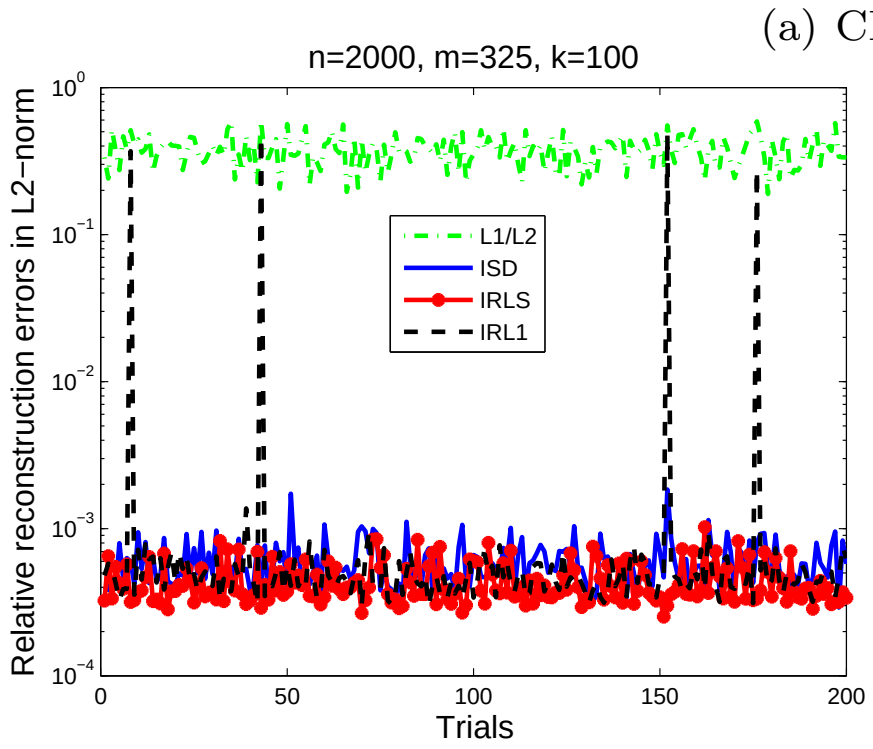

(b) Reconstruction errors in $\ell_2$ norm

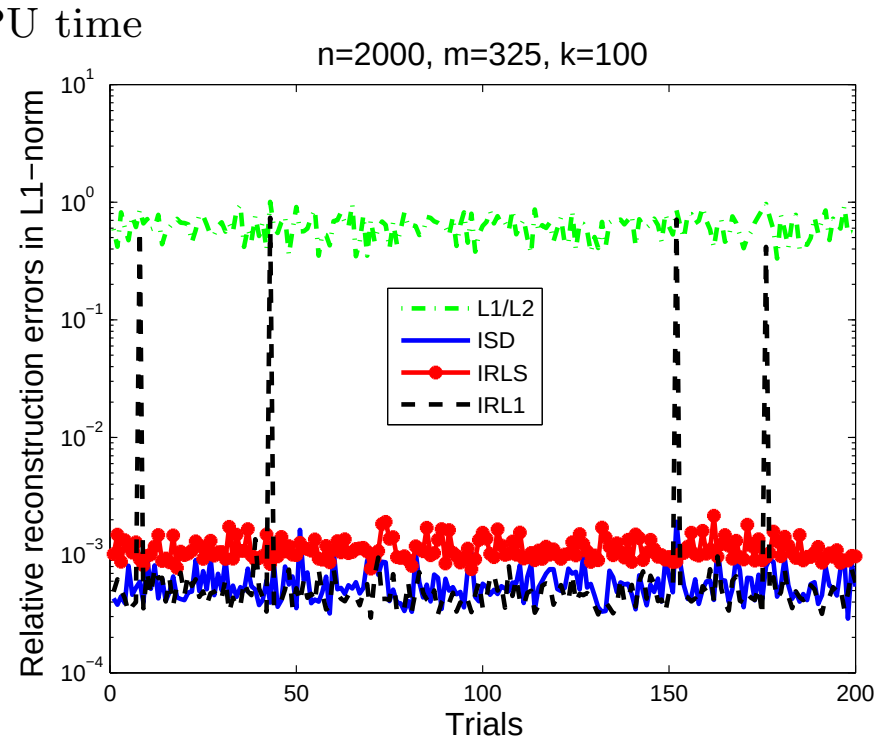

(c) Reconstruction errors in $\ell_1$ norm

Figure 7: Test set 3 with $\sigma = 0.0001$: Comparisons in CPU time and reconstruction errors

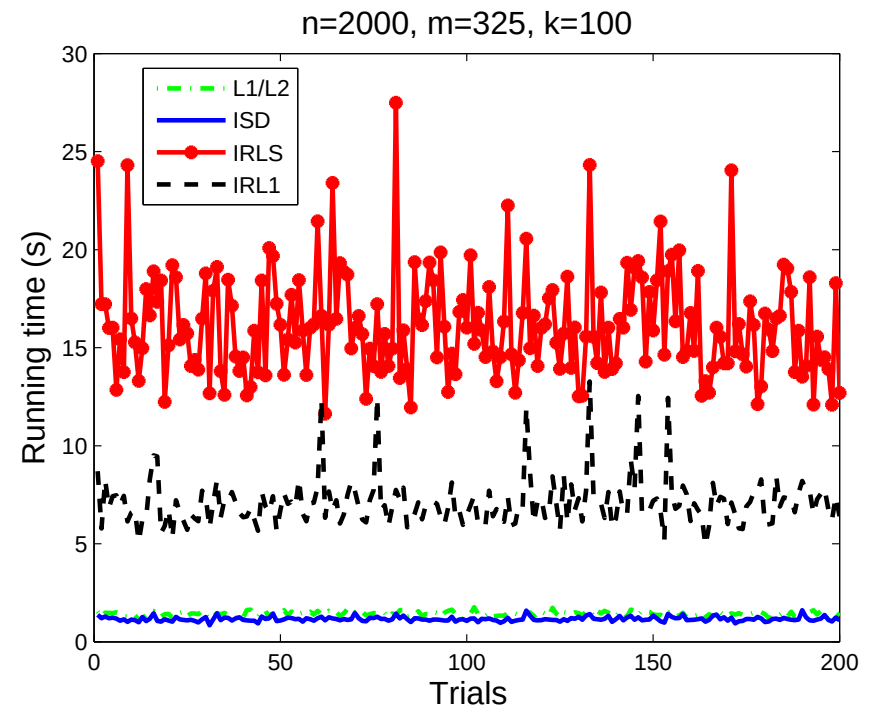

(a) CPU time

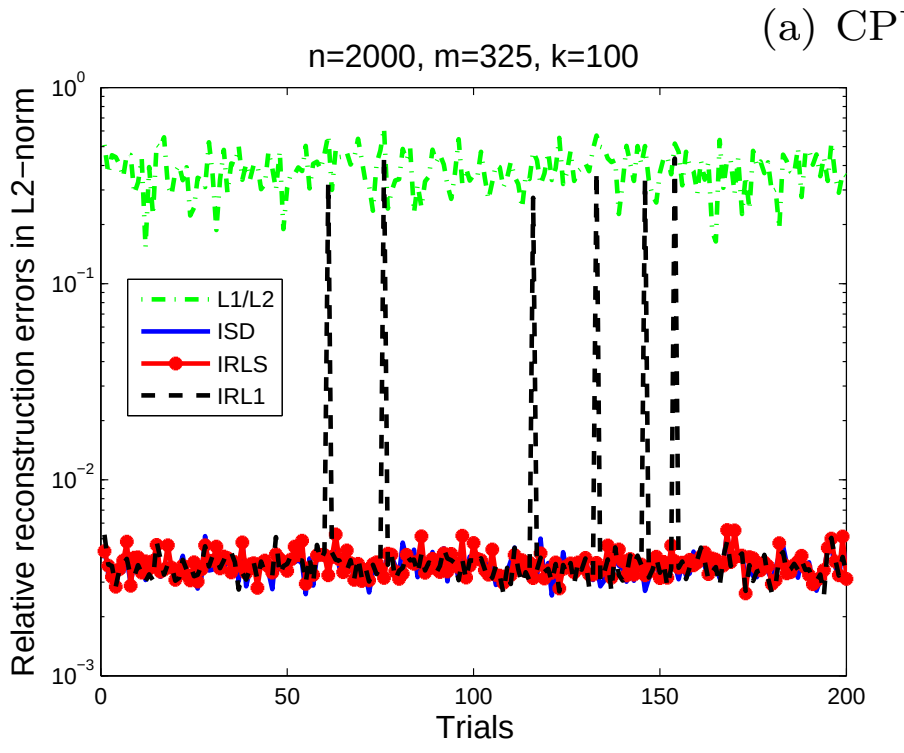

(b) Reconstruction errors in $\ell_2$ norm

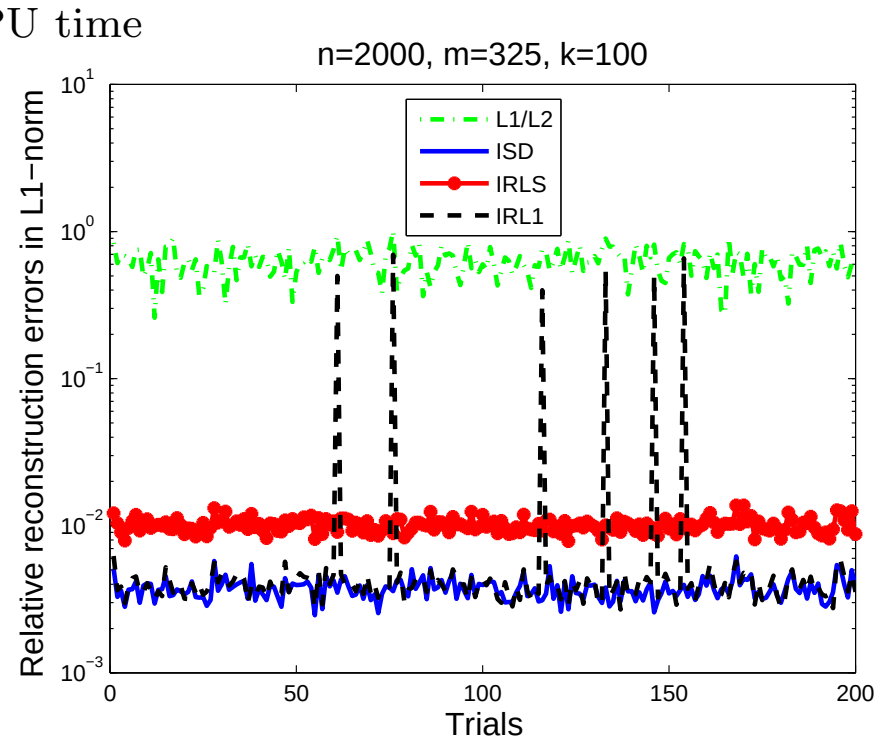

(c) Reconstruction errors in $\ell_1$ norm

Figure 8: Test set 3 with $\sigma = 0.001$: Comparisons in CPU time and reconstruction errors

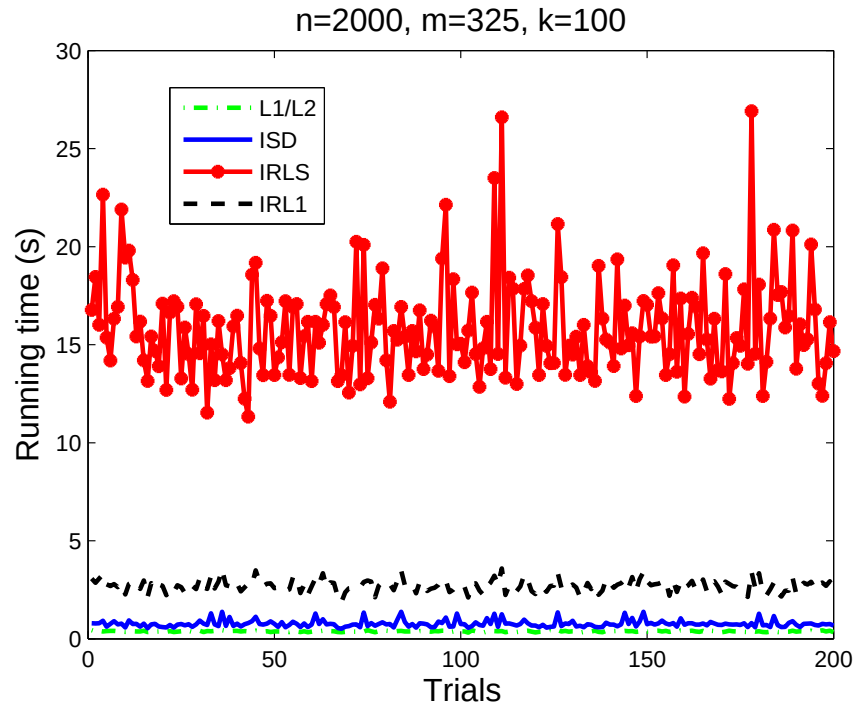

(a) CPU time

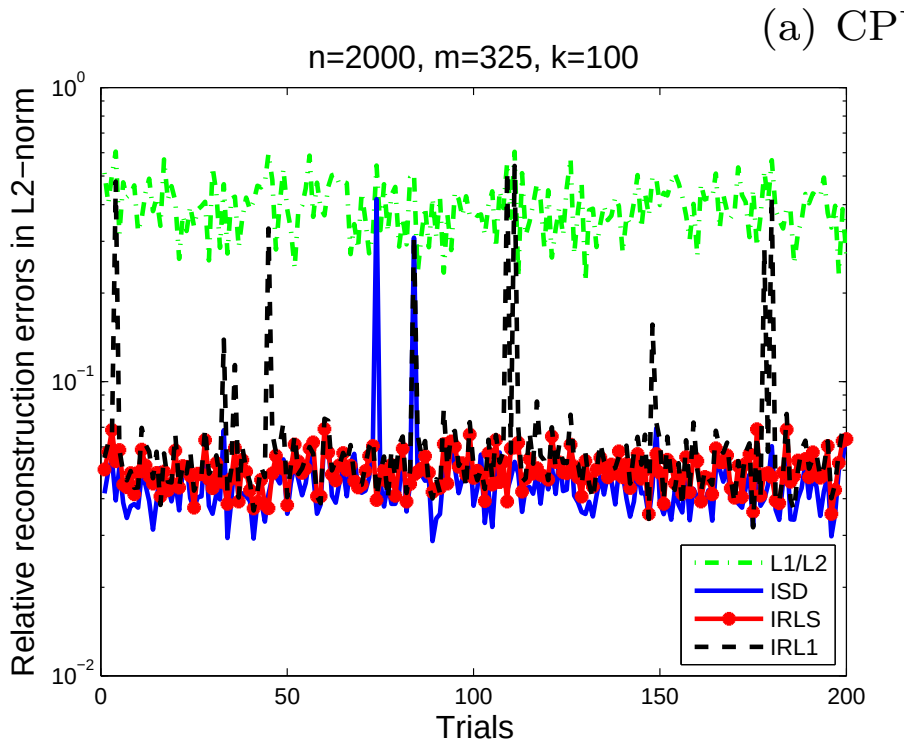

(b) Reconstruction errors in $\ell_2$ norm

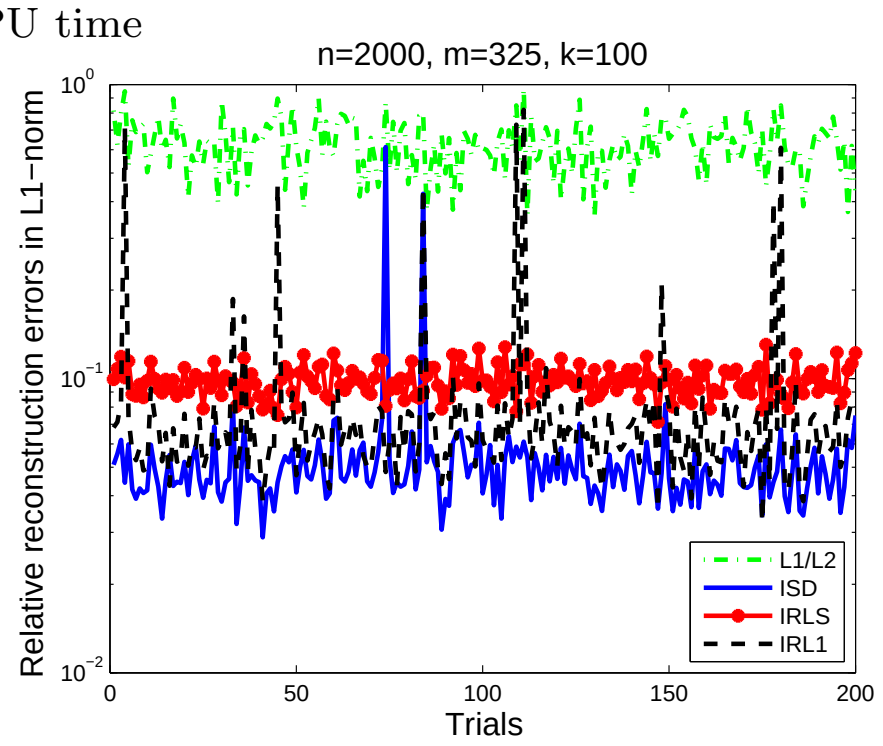

(c) Reconstruction errors in $\ell_1$ norm

Figure 9: Test set 3 with $\sigma = 0.01$: Comparisons in CPU time and reconstruction errors

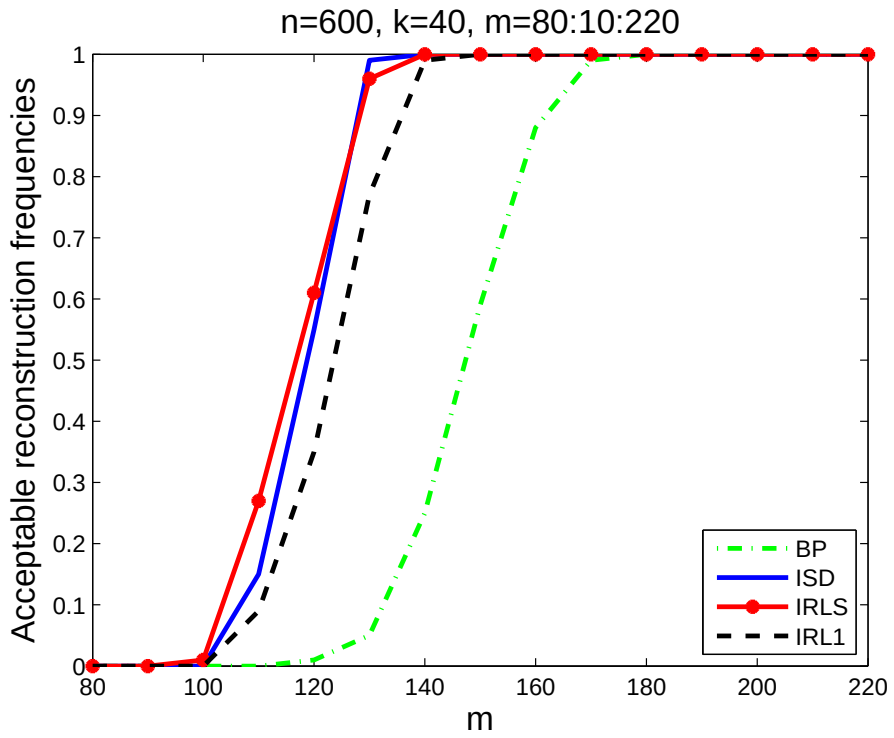

(a) $\lambda$=1: acceptable reconstruction frequencies

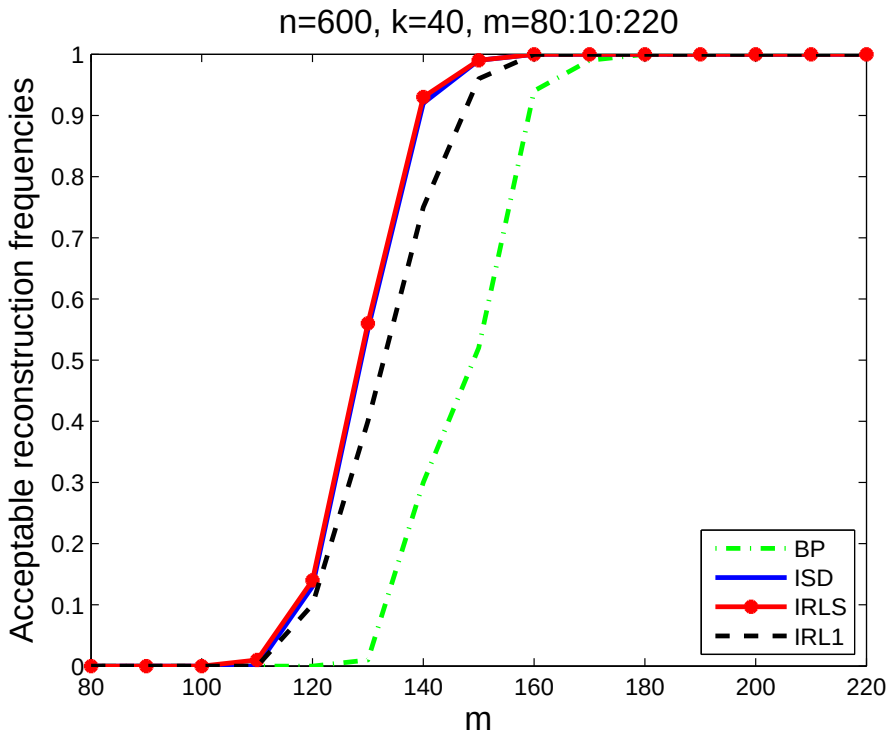

(b) $\lambda$=2: acceptable reconstruction frequencies

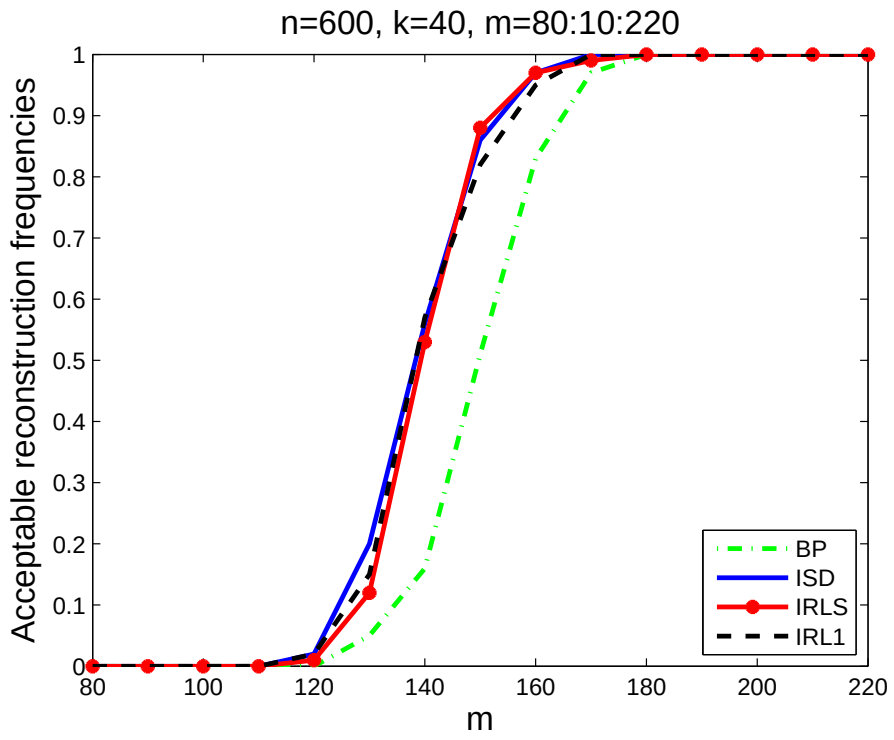

(c) $\lambda$=4: acceptable reconstruction frequencies

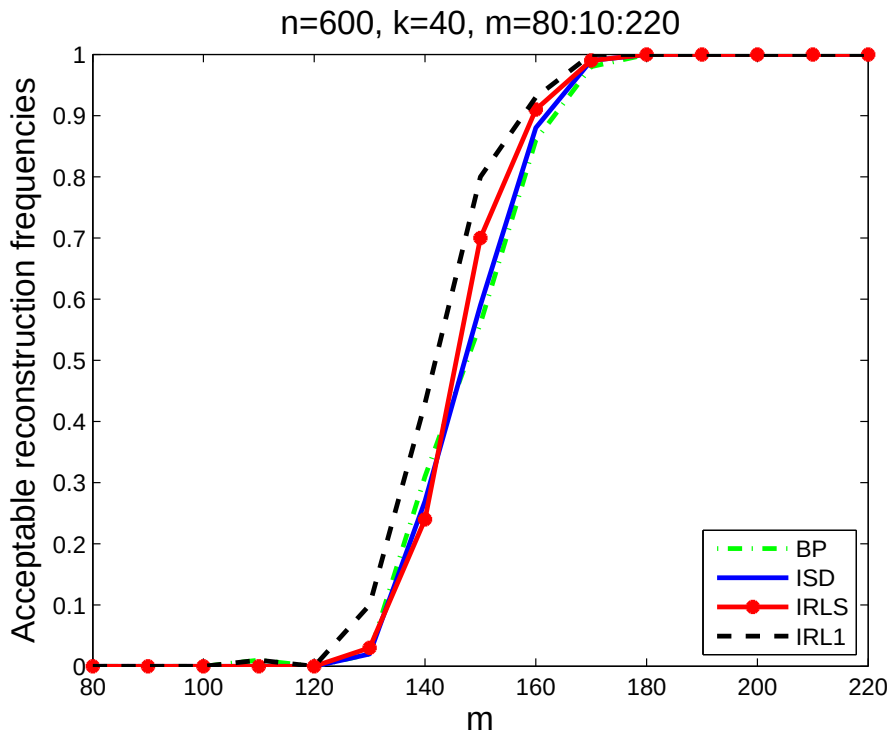

(d) Bernoulli: acceptable reconstruction frequencies

Figure 10: Test set 4 with sparse power–law decaying and Bernoulli signals: Comparisons in recoverability

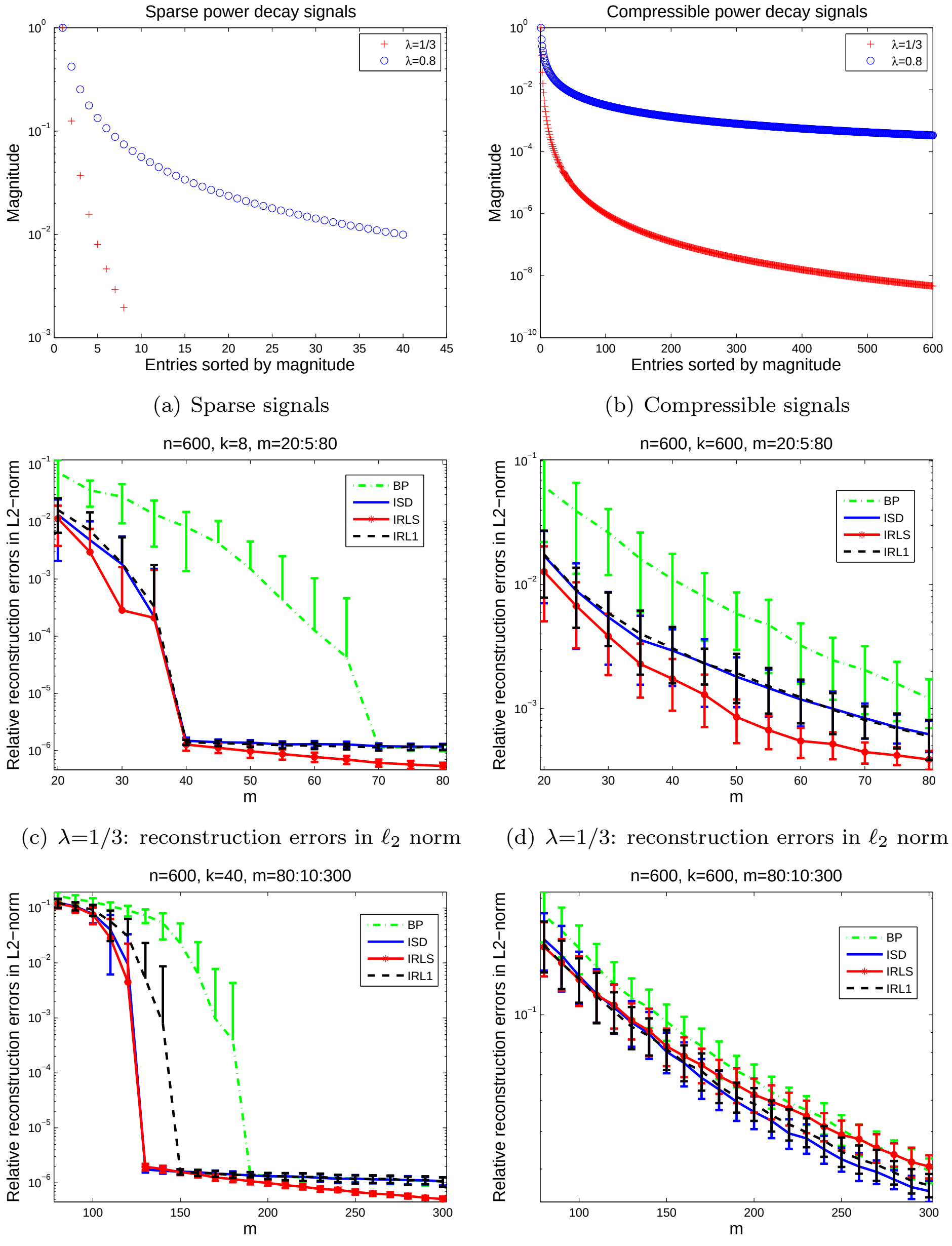

(a) Sparse signals

(b) Compressible signals

(c) $\lambda$=1/3: reconstruction errors in $\ell_2$ norm

(d) $\lambda$=1/3: reconstruction errors in $\ell_2$ norm

(e) $\lambda$=0.8: reconstruction errors in $\ell_2$ norm

(f) $\lambda$=0.8: reconstruction errors in $\ell_2$ norm

Figure 11: Test set 4 with sparse and compressible power–law decay signals. The sparse signals were obtained by removing the tails of the compressible signals. Comparisons in reconstruction errors. Fast decaying tails required for good performances of threshold–ISD, IRL1 and IRLS.

**Test Set** 5 : Figure 12 depicts the clean 128 × 128 Shepp–Logan phantom and its wavelet coefficients sorted by magnitude, which have a long tail and a sharp drop near 1600. Threshold–ISD, IRL1, and BP were tested to reconstruct the phantom from partial discrete cosine measurements both without and with white noise. The results are given in Figures 13 and 14, respectively. For Figure 13, a reconstruction $\tilde{x}$ was accepted if its 2–norm relative error $\|\tilde{x}(:) - \bar{x}(:)\|_2/\|\bar{x}(:)\|_2$ was within $10^{-3}$. Threshold–ISD was both faster than IRL1 and more accurate than IRL1 and BP in both of the tests.

Figure 15 presents the reconstructed phantoms from noisy measurements corresponding to $m = 3370$, in which subplots (b), (d) and (f) highlight the differences between the reconstructions and the clean phantom. Threshold–ISD gave a much higher signal–to–noise (SNR) ratio.

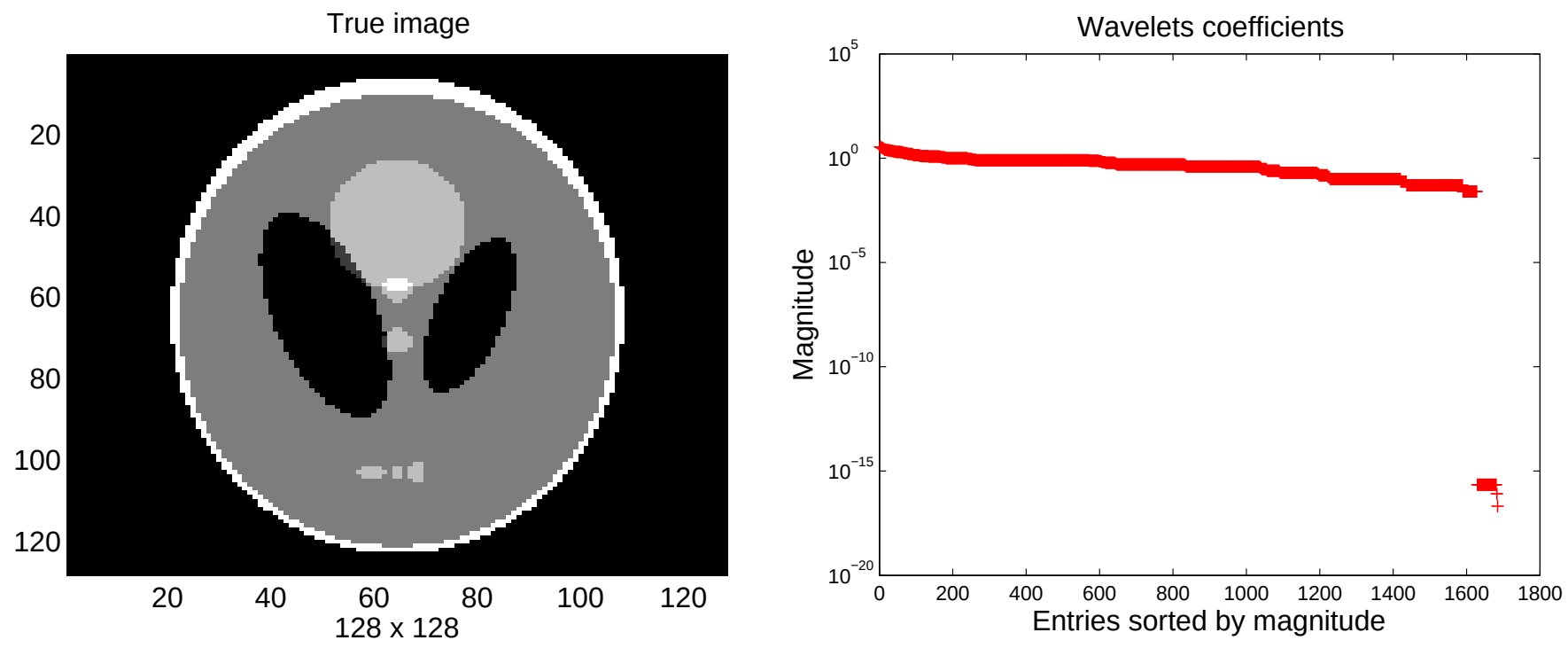

(a) (b)

Figure 12: Shepp–Logan phantom and its wavelets coefficients sorted by magnitude

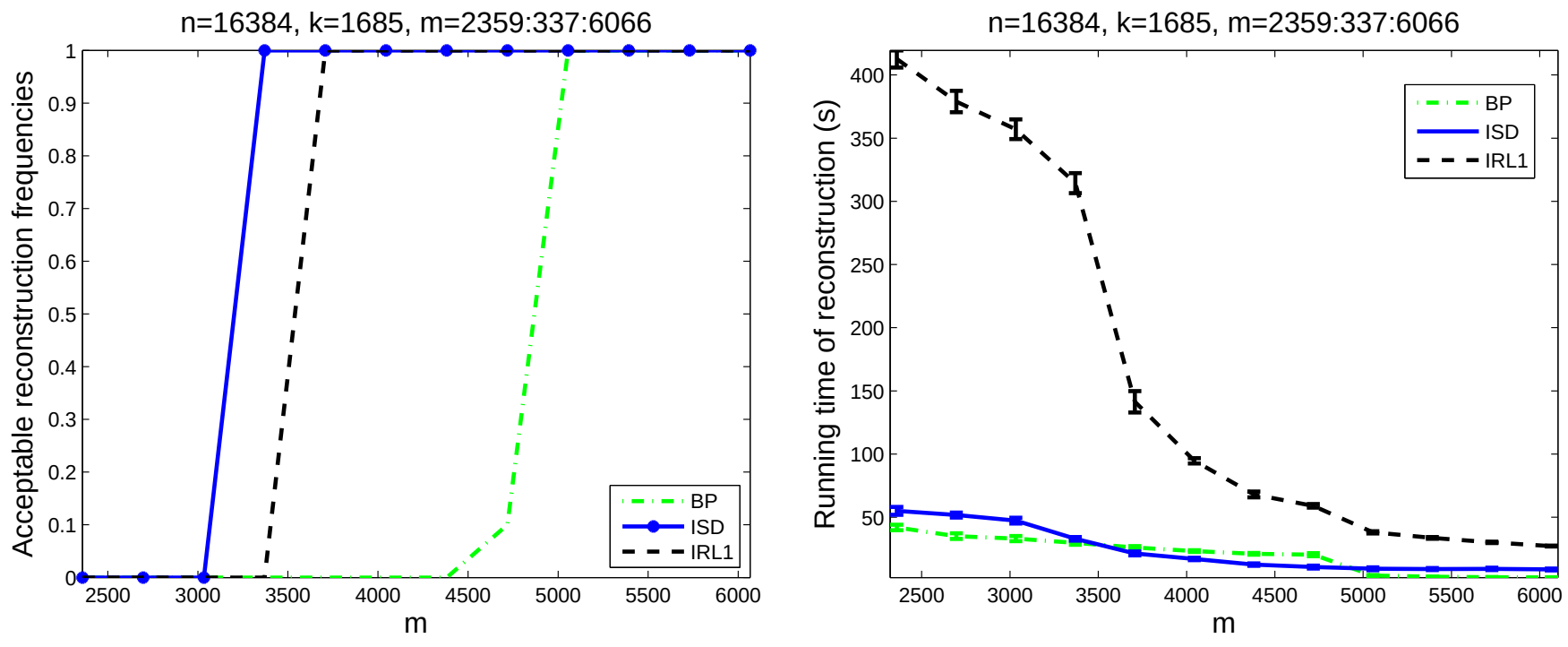

(a) (b)

Figure 13: Noiseless measurements: Acceptable reconstruction frequencies and running times

Figure 16 depicts the clean image "Cameraman" and its sorted wavelet coefficients, which form a long and slow decaying tail. As expected, threshold–ISD did not return a reconstruction significantly better than either BP or IRL1 as shown in Figure 17. Even though thresholding in the wavelet domain is not effective for natural images, we note that the authors of [19], however, have obtained medical images with much better quality by combining ISD (applied to total variation) with edge detection techniques that replace thresholding. For this and other reasons, we believe that ISD with effective support detection is potentially very powerful.

In summary, we compared threshold–ISD with IRLS, IRL1, and BP. Threshold–ISD can be solved as fast as BP yet achieves reconstruction quality as good as or better than the much slower IRLS. ISD relies on effective support detection. For threshold–ISD, its good performance requires fast–decaying signals.

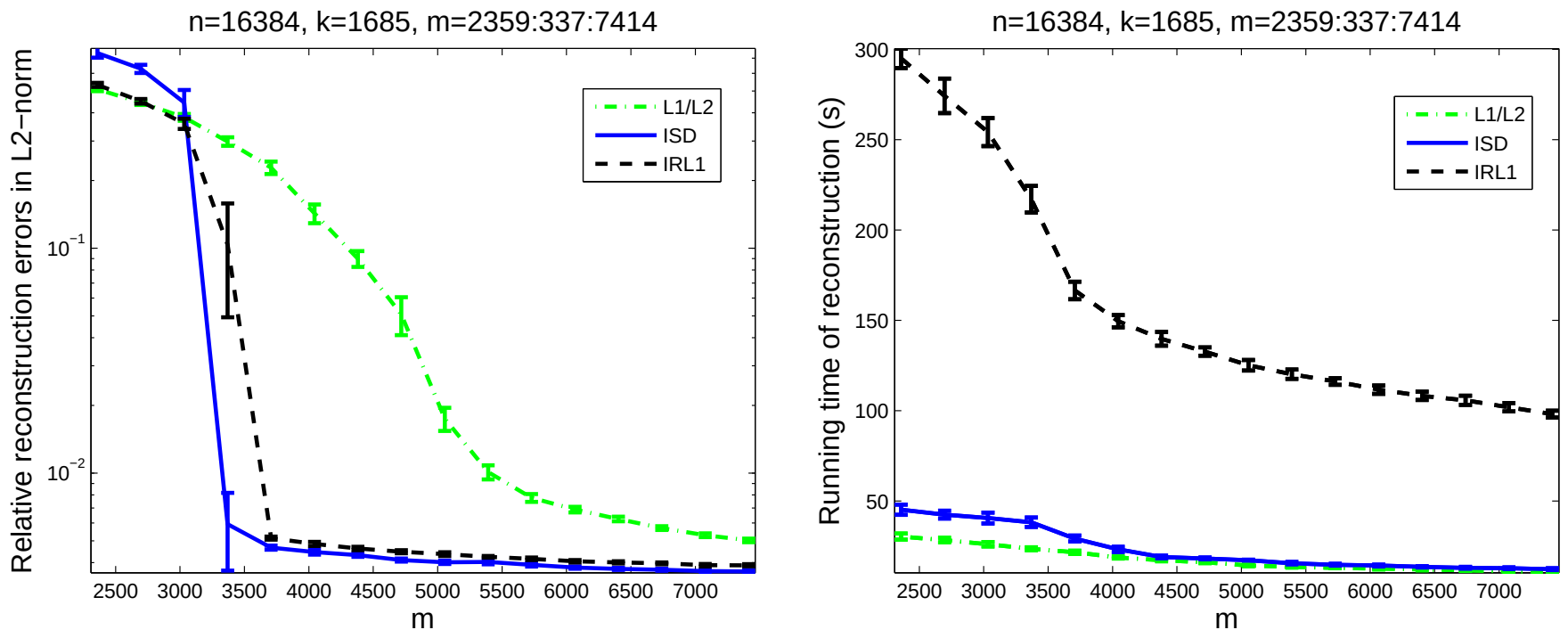

(a) (b)

Figure 14: Noisy measurements: Reconstruction errors and running times

# 6 Concluding Remarks

This paper introduces the iterative support detection method ISD for compressive sensing signal reconstruction. Both theoretical properties and practical performances are discussed. For signals with a fast decaying distribution of nonzeros, the implementation threshold–ISD equipped with the "first significant jump" thresholding rule is both fast and accurate compared to the classical approach BP and the state-of-the-art algorithms IRL1 and IRLS. Due to the limit of thresholding, threshold–ISD does not perform significantly better than its peers on other types of signals such as images.

However, support detection is not limited to thresholding. Effective support detection guarantees good performance of ISD. Therefore the future research includes studying specific signal classes and developing more effective support detection means, for example, by exploring signal structures (model–based CS [2]).

Since minimizing $\ell_1$ is not the only approach for compressive sensing signal reconstruction, another line of future exposition is to apply iterative support detection to other reconstruction approaches such as the greedy algorithms, Bayesian algorithms, dictionary–based algorithms, and many others. We also feel that the usefulness of the above "first significant jump" rule is not limited to threshold–ISD.

# Acknowledgement

The authors want to thank Professor Yin Zhang for making his YALL1 package available to us and Professors Weihong Guo and Yin Zhang for their valuable comments and suggestions that led to improvements of this report.

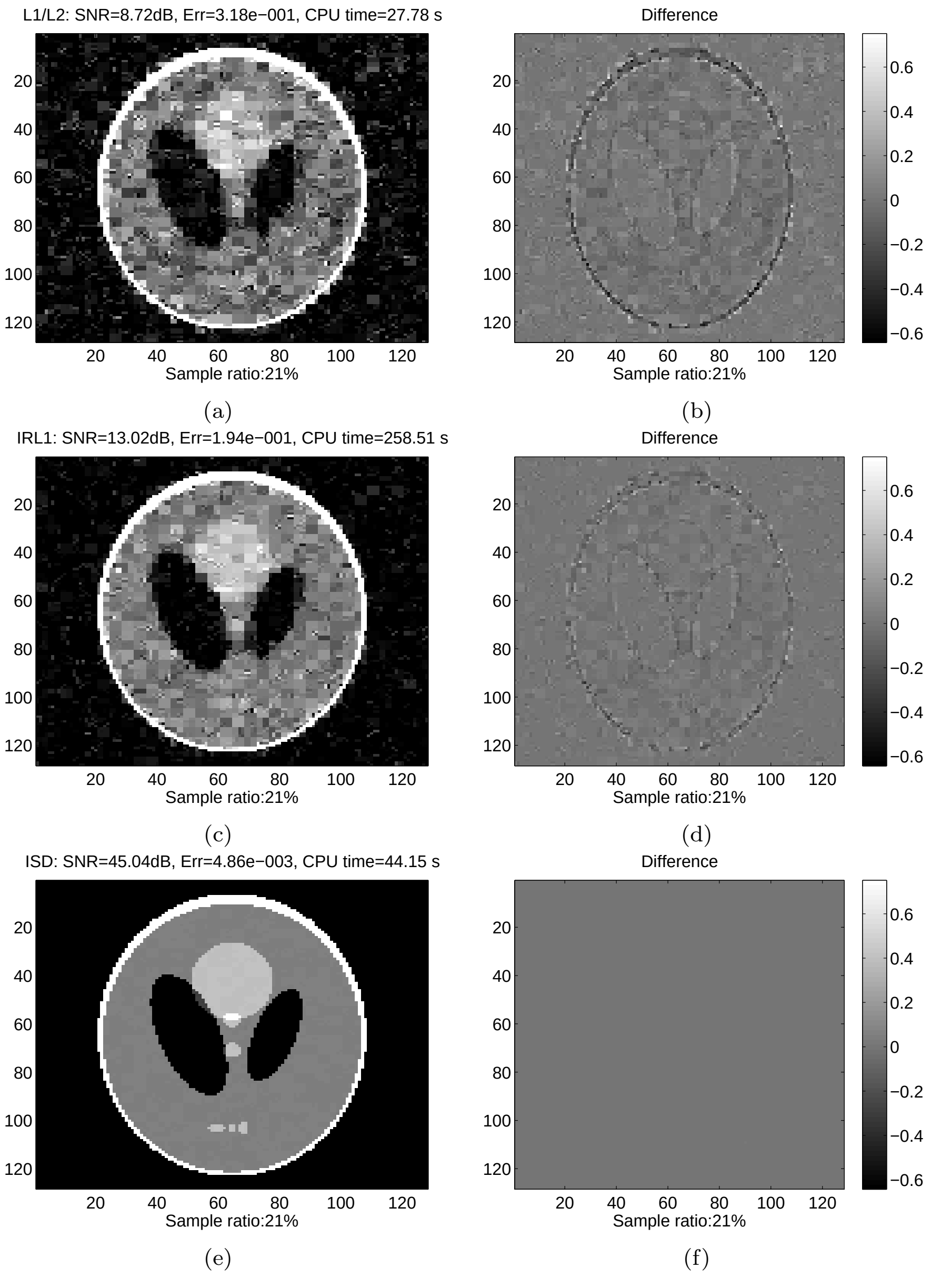

Figure 15: Noisy measurements: Reconstructed phantoms and highlighted errors corresponding to where $m = 3370$

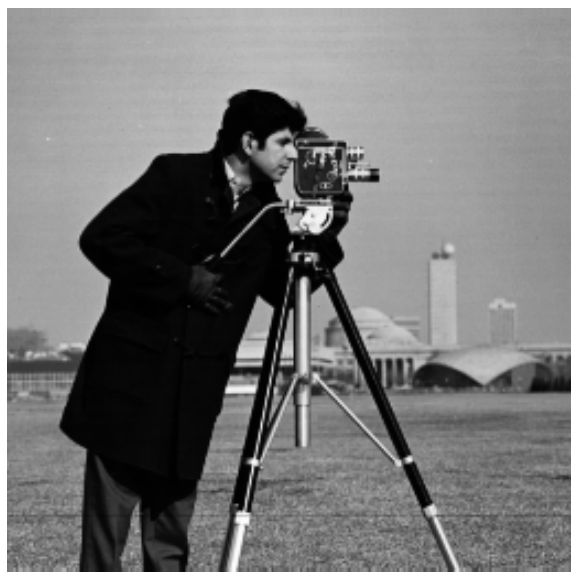

(a)

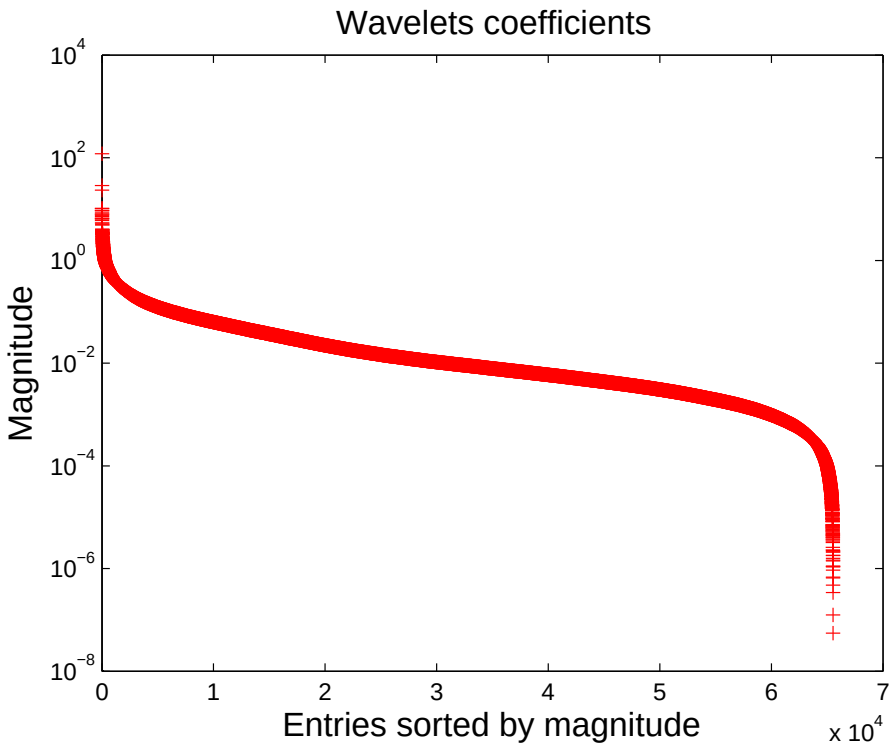

(b)

Figure 16:

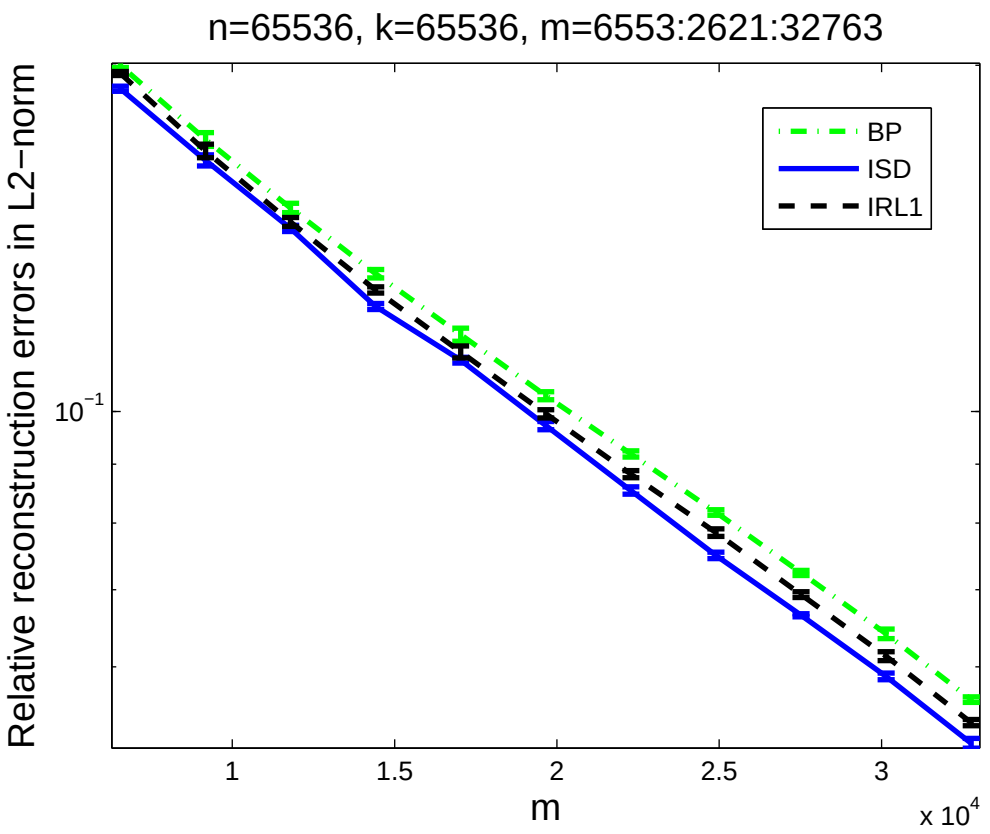

(a)

Figure 17: Reconstructing errors where the true signal is the wavelet coefficients of “Cameraman” and the sensing matrices are partial discrete cosine matrices